\pdfoutput=1

\documentclass[iop]{emulateapj}
\usepackage{epstopdf}
\usepackage{graphicx}
\usepackage{lineno}
\usepackage{footnote}

\shorttitle{Multifrequency WEBT Observations of S5~0716$+$714}
\shortauthors{Bhatta et al.}

\begin{document}

\title{Multifrequency Photo-polarimetric WEBT\footnotemark[\dag] Observation Campaign on the Blazar S5~0716$+$714: Source microvariability and Search for Characteristic Timescales}

\author{G.~Bhatta\altaffilmark{1},
{\L}.~Stawarz\altaffilmark{1},
M.~Ostrowski\altaffilmark{1},
A.~Markowitz\altaffilmark{2},
H.~Akitaya\altaffilmark{3},
A.~A.~Arkharov\altaffilmark{4}
R.~Bachev\altaffilmark{5},
E.~Ben\'itez\altaffilmark{6},
G.~A.~Borman\altaffilmark{7},
D.~Carosati\altaffilmark{8,9},
A.~D.~Cason\altaffilmark{10},
 R.~Chanishvili\altaffilmark{11},
G.~Damljanovic\altaffilmark{12}
S.~Dhalla\altaffilmark{13},
A.~Frasca\altaffilmark{14},
D.~ Hiriart\altaffilmark{15},
S-M.~Hu\altaffilmark{16},
R.~Itoh\altaffilmark{17},
D.~Jableka\altaffilmark{1},
S.~Jorstad\altaffilmark{18,19},
M.~D.~Jovanovic\altaffilmark{12},
K.~S.~Kawabata\altaffilmark{3},
S.~A.~Klimanov\altaffilmark{4},
O.~Kurtanidze\altaffilmark{11,20,21},
V.~M.~Larionov\altaffilmark{19,4},
D.~Laurence\altaffilmark{13},
G.~Leto\altaffilmark{14},
A.~P.~Marscher\altaffilmark{18},
J.~W.~Moody\altaffilmark{22},
Y.~Moritani\altaffilmark{23},
J.~M.~Ohlert\altaffilmark{24},
A.~Di~Paola \altaffilmark{25}
C.~M.~Raiteri \altaffilmark{26}
N.~Rizzi\altaffilmark{27},
A.~C.~Sadun\altaffilmark{28},
M.~Sasada\altaffilmark{18},
S.~Sergeev\altaffilmark{7},
A.~Strigachev\altaffilmark{5},
K.~Takaki\altaffilmark{17},
I.~S.~Troitsky\altaffilmark{19},
T.~Ui\altaffilmark{17},
M.~Villata\altaffilmark{26},
O.~Vince\altaffilmark{12},
J.~R.~Webb\altaffilmark{13},
M.~Yoshida\altaffilmark{3},
and S.~Zola\altaffilmark{1,29}
}

\altaffiltext{1}{Astronomical Observatory of Jagiellonian University, ul. Orla 171, 30-244 Krakow, Poland}
\altaffiltext{2} {Center for Astrophysics \& Space Sciences, University of California, San Diego, 9500 Gilman Dr., La Jolla, CA 92093-0424, USA}
\altaffiltext{3}{Hiroshima Astrophysical Science Center, Hiroshima University, Higashi-Hiroshima, Hiroshima 739-8526, Japan}
\altaffiltext{4} {Main (Pulkovo) Astronomical Observatory of RAS, Pulkovskoye shosse, 60, 196140 St. Petersburg, Russia}
\altaffiltext{5}{Institute of Astronomy, Bulgarian Academy of Sciences, 72, Tsarigradsko Shosse Blvd., 1784 Sofia Bulgaria}
\altaffiltext{6}{Instituto de Astronom\'ia, Universidad Nacional Aut\'onoma de M\'exico, Mexico DF, Mexico}
\altaffiltext{7}{Crimean Astrophysical Observatory, P/O Nauchny, Crimea, 298409, Russia}
\altaffiltext{8} {EPT Observatories, Tijarafe, La Palma, Spain}
\altaffiltext{9} {INAF, TNG Fundacion Galileo Galilei, La Palma, Spain}
\altaffiltext{10} {Private address, 105 Glen Pine Trail, Dawnsonville, GA 30534, USA}

\altaffiltext{11} {Abastumani Observatory, Mt. Kanobili, 0301 Abastumani, Georgia}

\altaffiltext{12} {Astronomical Observatory, Volgina 7, 11060 Belgrade, Serbia}
\altaffiltext{13} {Florida International University, Miami, FL 33199, USA}
\altaffiltext{14} {INAF - Osservatorio Astrofisico di Catania, Italy}
\altaffiltext{15}{Instituto de Astronom\'ia, Universidad Nacional Aut\'onoma de M\'exico, Ensenada, Mexico}
\altaffiltext{16} {Shandong Provincial Key Laboratory of Optical Astronomy and Solar-Terrestrial Environment, Institute of Space Sciences, Shandong University at Weihai, 264209 Weihai, China}
\altaffiltext{17}{Department of Physical Science, Hiroshima University, Higashi-Hiroshima, Hiroshima 739-8526, Japan}
\altaffiltext{18} {Institute for Astrophysical Research, Boston University, 725 Commonwealth Avenue, Boston, MA 02215, USA}
\altaffiltext{19} {Astronomical Institute, St. Petersburg State University, Universitetskij Pr. 28, Petrodvorets, 198504 St. Petersburg, Russia}

\altaffiltext{20} {Engelhardt Astronomical Observatory, Kazan Federal University, Tatarstan, Russia}
\altaffiltext{21} {Landessternwarte Heidelberg-Konigstuhl, Germany}

\altaffiltext{22} {Physics and Astronomy Department, Brigham Young University, N283 ESC, Provo, UT, USA 84602}
\altaffiltext{23}{Kavli Institute for the Physics and Mathematics of the Universe (KavliI PMU), The University of Tokyo, 5-1-5 Kashiwa-no-Ha, Kashiwa City Chiba, 277-8583, Japan}
\altaffiltext{24} {Astronomie Stiftung Tebur, Fichtenstrasse 7, 65468 Trebur, Germany}
\altaffiltext{25}{INAF - Osservatorio Astronomico di Roma, via Frascati 33, 00040 Monte Porzio, Italy}
\altaffiltext{26}{INAF - Osservatorio Astrofisico di Torino, Italy}
\altaffiltext{27} {Sirio Astronomical Observatory Castellana Grotte, Italy}
\altaffiltext{28} {Department of Physics, Univ. of Colorado Denver, CO, USA }
\altaffiltext{29}{Mt. Suhora Observatory, Pedagogical University, ul. Podchorazych 2, 30-084 Krakow, Poland}

\email{email: {\tt gopalbhatta716@gmail.com}}

\footnotetext[\dag]{The data collected by the WEBT Collaboration are stored in the WEBT archive; for questions regarding their availability, please contact the WEBT President Massimo Villata (\texttt{villata@oato.inaf.it).}} 

\begin{abstract}
Here we report on the results of the WEBT photo-polarimetric campaign targeting the blazar S5~0716+71, organized in March 2014 to monitor the source simultaneously in BVRI and near IR filters. The campaign resulted in an unprecedented dataset spanning $\sim 110$\,h of nearly continuous, multi-band observations, including two sets of densely sampled polarimetric data mainly in R filter. During the campaign, the source displayed pronounced variability with peak-to-peak variations of about $30\%$ and ``bluer-when-brighter'' spectral evolution, consisting of a day-timescale modulation with superimposed hourlong microflares characterized by $\sim 0.1$\,mag flux changes. We performed an in-depth search for quasi-periodicities in the source light curve; hints for the presence of oscillations on timescales of $\sim 3$\,h and $\sim 5$\,h do not represent highly significant departures from a pure red-noise power spectrum. We observed that, at a certain configuration of the optical polarization angle relative to the positional angle of the innermost radio jet in the source, changes in the polarization degree led the total flux variability by about 2\,h; meanwhile, when the relative configuration of the polarization and jet angles altered, no such lag could be noted. The microflaring events, when analyzed as separate pulse emission components, were found to be characterized by a very high polarization degree ($> 30\%$) and polarization angles which differed substantially from the polarization angle of the underlying background component, or from the radio jet positional angle. We discuss the results  in the general context of blazar emission and energy dissipation models.\end{abstract}

\keywords{acceleration of particles --- polarization --- radiation mechanisms: non-thermal --- galaxies: active --- BL Lacertae objects: individual (S5\,0716+714) --- galaxies: jets}

\section{Introduction \label{sec:intro}}

Blazars, a subclass of radio-loud active galactic nuclei (AGN), are usually identified by their Doppler-boosted non-thermal emission across the entire electromagnetic spectrum, originating from relativistic jets aligned near the line of sight \citep[e.g.,][]{Meier12}. They exhibit significant, often dramatic variability at different wavelengths and on diverse timescales, ranging from minutes up to years and decades. In particular, flux fluctuations by a few percent observed on timescales of minutes and hours, are usually termed as an intraday/intranight variability (IDV/INV), or a microvariability \citep{WNW95}. Blazar microvariability at various frequencies has been studied by a number of authors since the late 70s, and was initially thought to result from the instrumental artifacts or external causes \citep[environmental scintillation, gravitational micro-lensing, etc.; see, e.g.,][]{Schneider87,Melrose94}. Later, however, with the improvement of sensitive instruments such as charged coupled device (CCD) cameras, and polarimetric measurements, those rapid and small-amplitude brightness fluctuations were fairly proved to be source-intrinsic, and in addition to originate in the innermost parts of relativistic jets \citep[e.g.,][]{Pollock07,Sasada08,Goyal12}. Since the blazar optical emission zone is not spatially resolved on (sub)-milliarc-second scales by any currently operating telescopes, the study of microvariabilty can be therefore used to understand the structure of AGN outflows close to/at the jet base, and to constrain the main physical processes operating therein that shape the production of high-energy particles and non-thermal emission of blazar sources. Yet, despite a substantial observational effort, as well as a comprehensive theoretical discussion on the topic, with various models and scenarios proposed, blazar variability (and microvariability in particular) is still relatively poorly understood.

The polarimetric blazar variability in the optical band has been subjected to an extensive investigation in the past. The temporal polarization changes, observed on timescales from minutes to years, in most of the cases appear random, with no obvious or only a weak correlation between the polarization degree and the total flux \citep[e.g.,][]{Hagen-Thorn80,Moore82,Tommasi01,Cellone07,Ikejiri11,Itoh13,Gaur14,Raiteri13}. Only in some particular sources during certain periods the polarized and total fluxes have been shown to vary in accord \citep[e.g.,][]{Tosti98,Hagen-Thorn08,Agudo11,Sorcia13,Bhatta15}. Also, more recently, several cases of prominent swings/rotations in the optical polarization angle accompanying high-energy $\gamma$-ray outbursts of the brightest blazars have been reported \citep{Abdo10,Jorstad10,mar08,mar10,Larionov13,Blinov15}. These results imply all together a complex magnetic field structure that determines the observed properties of the blazar synchrotron emission at optical wavelengths, including both the large-scale uniform component (often modeled in terms of a `grand-design' helix), and also a smaller-scale turbulent component (eventually only partly organized by the passage of shock waves and/or velocity shear within the outflow).

S5~0716+714 is one of the best known BL Lac objects, at a redshift of approximately $z = 0.31 \pm 0.08$ \citep[see][]{Nil08,danforth13}, classified as an `Intermediate Synchrotron Peaked' (IBL) blazar based on the location of its synchrotron peak in the $\nu F_{\nu}-\nu$ representation around frequencies of $\sim 10^{14}-10^{15}$\,Hz \citep{Ackermann11}. Since its discovery in 1979 by \citet{Kuhr81}, it has been the subject for numerous studies across all the available electromagnetic spectrum, due to its brightness, high declination in the sky, and its never ceasing variability with almost 100\% duty cycle \citep[e.g.][]{Heidt96}. At radio frequencies, S5~0716+714 appears on milliarc-second scales as a flat-spectrum, IDV, and superluminal source, characterized by apparent velocities of various jet features reaching $37 c$ \citep{Bach05,Jorstad01,Rani15}, and a very high brightness temperature of the compact core \citep{Ostorero06}. The X-ray emission continuum of the blazar is in general concave, marking the transition from the synchrotron to the inverse-Compton emission components in the observed spectrum \citep{Ferrero06,Foschini06}. S5~0716+714 has been also detected at $\gamma$-ray photon energies by the EGRET, AGILE, and {\it Fermi}-LAT satellites \citep[see, e.g.,][and references therein]{Ghisellini97,Villata08,Rani13,Liao14}, as well as by the MAGIC Cherenkov telescope \citep{Anderhub09}.

At optical frequencies, S5~0716+714 appears as a bright, highly polarized, and highly variable source. Long-term optical light curves of the blazar are presented in \citet{nesci05} and \citet{rait03}, and its general optical polarization properties are discussed in \citet{Impey00} and \citet{Ikejiri11}. It was shown repeatedly that optical flux changes of S5~0716+714 do not correlate with radio variability \citep{rait03,Ostorero06}, but instead with $\gamma$-ray flares \citep[e.g.,][]{Villata08,Rani13,Liao14}, flares which in addition seem to be accompanied by large swings in the optical polarization angle \citep{Larionov13,Chandra15}. Quasi-periodicity has been claimed in the optical light curves of the source for different epochs and at various timescales of hours, days, and years \citep{rait03,Gupta08,Gupta09,Gupta12}. The optical microvariability of S5~0716+714 has been widely investigated by a number of authors, who found high or very high INV duty cycle, often (though not always) bluer-when-brighter spectral behavior, red noise-type power spectra, and in some cases clear polarization degree--flux correlations \citep{Nesci02,mont06,Sasada08,Stalin09,Poon09,Carini11,Chandra11,Wu12,Zhang12,Dai13,Hu14,Bhatta15,Agarwal16}.

\begin{table*}[t!]
\caption{Observatories Contributing to the 2014 WEBT Campaign on S5~0716+714}
\label{tab:obslog}
\centering        
\begin{tabular}{ l l l l l}  
\hline\hline         
No.	&	Observatory	&	Telescope&Filter (PH)  & Filter (PL)\\    
\hline                  
1	&	Abastumani Obs., Georgia	&	70cm	&BVRI		&---\\
2	&	Astronomical Obs., Krak\'ow, Poland	&	50cm	&BVRI		&---\\ 
3	&Astronomical Station Vidojevica, Serbia &	60cm	&	BVRI	&---\\
4	&	Belogradchik, Bulgaria	&	60cm	&	BVRI	&---\\
5	&     Crimean Astrophysical Obs., Russia&	70cm	&	BVRI	&R\\
6	&	Campo Imperatore, Italy &	110cm	&	JHK	&---\\
7	&EPT Observatories Tijarafe La Palma� Spain &40cm Ritchey Chretien&R&---\\
8	&	Fairborn, Arizona, USA &	APT 80cm	&	BVRI	&---\\
9	&	Higashi-Hiroshima, Kanata, Japan & 150cm	&BVRI		&R\\
10	&	L'Ampolla, Spain	&	 36cm	&	BVRI	&---\\
11	&	Lowell Obs., Perkins, Flagstaff, AZ, USA & 180cm	&	BVRI	&BVRI\\
12	&	Michael Adrian Obs., Germany	&	120cm	&BVRI		&---\\
13	&Astronomical Obs. Sirio Castellana Grotte, Italy &	25cm	&R &---\\
14	&	SARA/Kitt Peak, USA	&	90cm	&	BVRI	&---\\	
15	&	St. Petersburg University, Russia &	40cm	&	BVRI	&WL \\
16	&	Suhora Observatory, Poland&	90cm	&BVRI		&---\\
17	&	T-11 Mayhill, New Mexico, USA&	51cm	&BVRI		&---\\
18	&	T-21 Mayhill, New Mexico, USA&	43cm	&BVRI		&---\\
19	&	T-24 Auberry, California, USA&	61cm	&VI		&---\\
20	&	Weihai Obs. of Shandong Univ., China&	100cm	& BVRI		&\\
\hline          
\end{tabular}

\vspace{0.25cm}
PH $\rightarrow$ Photometric; PL  $\rightarrow$ Polarimetric; WL $\rightarrow$ White Light
\end{table*}

Here we present the result of the multifrequency photometric and polarimetric monitoring campaign on S5~0716+714 through the Whole Earth Blazar Telescope (WEBT), which took place from March 2nd to 6th, 2014 (see \S\,\ref{sec:obs}). The main objective of the campaign was to monitor the source continuously for an extended period of time, to study its variations in flux, color, polarization degree (PD), and polarization angle (PA) simultaneously and with unprecedented details, building upon the previously undertaken successful WEBT monitoring campaigns targeting the blazar (by \citealt{Villata00} in Feb 16--19, 1999, \citealt{Ostorero06} in November 6--20, 2003, and \citealt{Bhatta13} in February 22--25, 2009). With the given duration of the campaign and its extremely dense, minute-scale sampling of the source light curve, the data could be subjected to a meaningful and robust time series analysis, in search of temporal characteristics (including possible periodicity) on timescales from a few hours to a day (\S\,\ref{sec:analysis}), i.e. the timescales which are basically unconstrained in either intra-night observations conducted by a single ground-based telescope, or typical long-term monitoring programs consisting of individual exposures isolated by days and weeks. The gathered rich dataset constrains uniquely the physics of the emission zone in S5~0716+714, and blazar emission models in general (\S\,\ref{sec:con}).

\section{Observations}
\label{sec:obs}

The WEBT\footnote{\texttt{\textbf{http://www.oato.inaf.it/blazars/webt/}}} multifrequency photometric and polarimetric monitoring campaign on S5~0716+714 was originally scheduled for March 3rd and 4th, 2014, but due to an extraordinary participation of the observers all around the globe, it had been extended to five days. All in all, 26 observers from 20 observatories monitored the source in various photo-polarimetric filters from March 2nd to 6th, 2014. During the campaign, the weather, on most of the telescope sites, was photometric enough to allow for a fair amount of multifrequency variability data. Hence the campaign resulted in photometric data in B, V, R, and I bands nearly continuous for five days, polarimetric data mainly in R filter for two days, and some near infrared data in J, H and K filters for few hours.

To achieve consistency and homogeneity over exposures of multiple observation sites and the instruments, a common set of instructions was followed by the observers. In particular, the same set of comparison stars 3, 4, and 6 from \citet{vill98} was used for the photometry. The participating observers carried out photometry for their images using a common set of standard procedures before they provided the data, containing instrumental magnitudes and the uncertainties of the source and the comparison stars in magnitudes, for the final compilation. Table\,\ref{tab:obslog} lists the names of the participating observatories along with their locations, telescope sizes, and filters used. 

\begin{figure*}[t!]
\centering
\includegraphics[width=\textwidth]{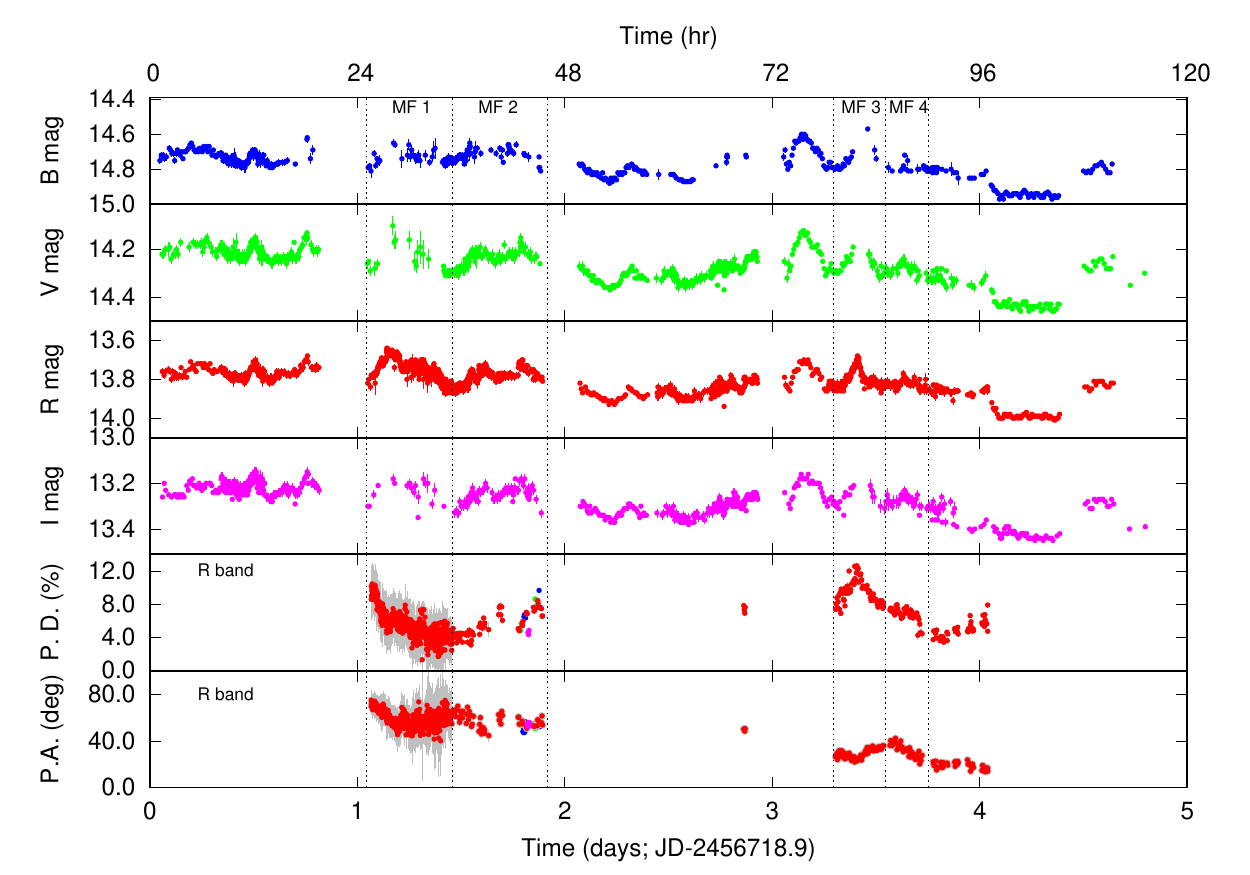}
\caption{The light curves of S5~0716+714 corresponding to all the data gathered during the 2014 WEBT campaign. In the upper panel, filters B, V, R, I, are presented by blue, green, red, magenta, respectively. In the lower panels, PD (middle) and PA (bottom) in B (blue), V (green), R  (red) and I (magenta) filters are shown. The dotted vertical lines mark the four microflares with polarimetric coverage analyzed in more detail in \S\,\ref{sec:mod}, and labeled as MF1 and MF2, MF3 and MF4.}
\label{rawlightcurve}
\end{figure*}

\begin{figure}[bh!]
\centering
\includegraphics[width=\columnwidth]{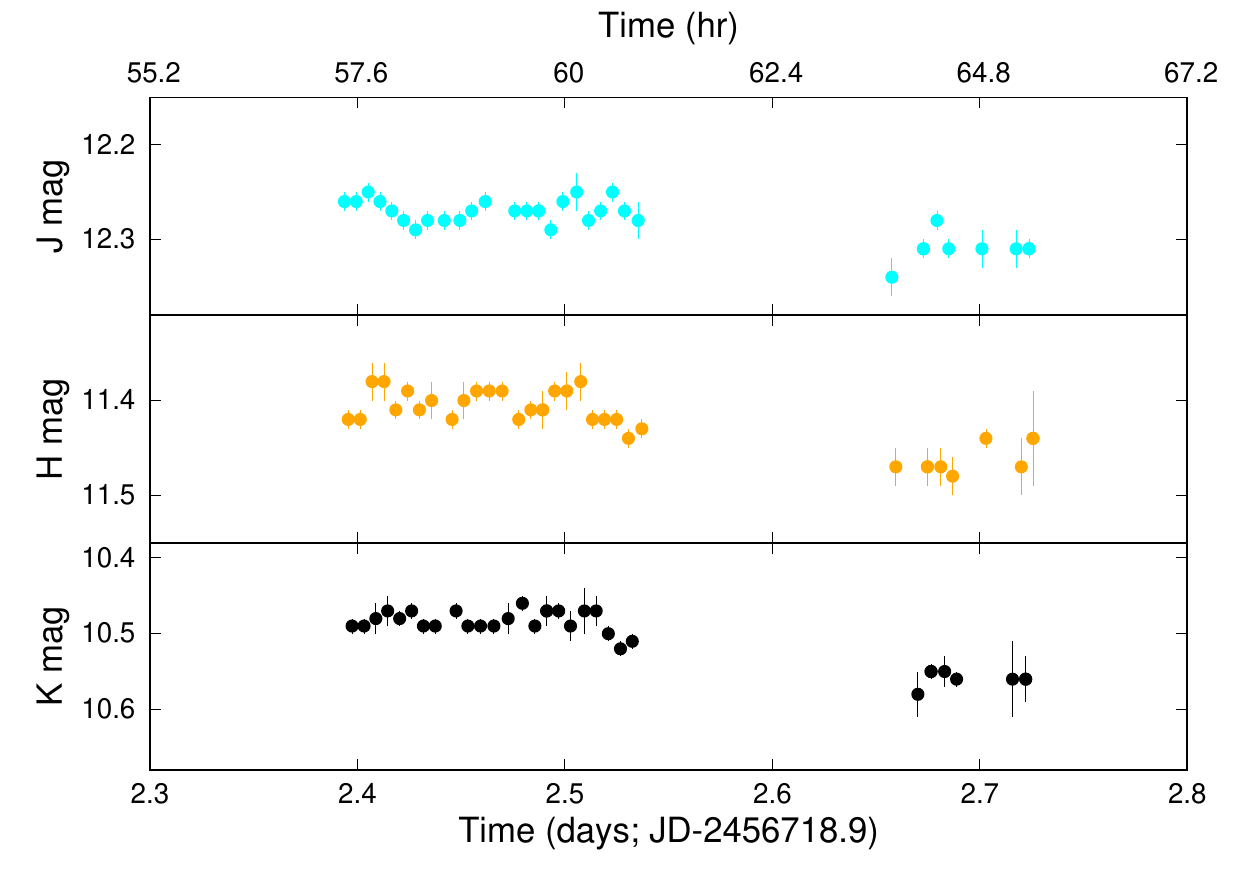}
\caption{The NIR light curves of S5~0716+714 from the 2014 WEBT campaign in filters J (cyan; top panel), H (yellow; middle panel), and K (black; bottom panel).}
\label{NIR}
\end{figure}

Standard procedures for aperture photometry have been used to extract magnitudes and related uncertainties from the scientific images after bias, dark, and flat-field corrections. Apertures of about 2-4 arcseconds, the corresponding number of pixels depending upon the instrument and the camera,  were chosen so as to have minimum scatter in the comparison stars in the same field. From the data collected by various observers, magnitudes with uncertainties less that $4\%$ were selected for the final compilation. Besides, data exhibiting sudden large jumps from the previous data points were also analyzed carefully before they were included in the analysis. The amount of data that were excluded from the final analysis contribute less than $3\%$ of the total data gathered during the whole campaign. Thus the number of photometric data points included in the final analysis are 548, 776, 1921 and 723 in the filters B, V, R and I, respectively. The obtained optical light curves in these filters  are presented in Figure\,\ref{rawlightcurve}. The accompanying much shorter NIR light curves of S5~0716+71 from the 2014 WEBT campaign in filters J, H, and K, are presented in Figure\,\ref{NIR}.

Unlike the photometric data provided by all the involved observatories, the polarimetric data were mainly obtained with the 70 cm AZT-8 reflector of the Crimean Astrophysical Observatory, the 40 cm LX-200 telescope in St. Petersburg, the 1.8 m Perkins telescope of Lowell Observatory, and the Kanata 1.5 m telescope equipped with HOWPol. The telescopes in Crimea and St. Petersburg use photo-polarimeters based on ST-7 CCDs, whereas Lowell Observatory uses the PRISM camera. For the details on these instruments and the methods the readers are directed the following references: \citet{Larionov13} for AZT-8 reflector and  LX-200 telescope, \citet{Jorstad10} for Perkins telescope, and \citet{Kawabata08} for Kanata HOWPol.

\section{Analysis and Results}
\label{sec:analysis}

The gathered photometric data are nearly continuous over the five-day campaign, however continuously sampled polarimetric data could be collected only in two one-day sets separated by a day. Therefore, the analysis is carried out in two parts. The first part includes the analysis of photometric data only, and the second part consists of the analysis of the data involving all the photometric and polarimetric data available. The analysis focusing on characteristic variability timescales and correlations between different fluxes in photo-polarimetric bands is presented in the following sections.

\begin{figure*}[t!]
\centering
\includegraphics[width=\textwidth]{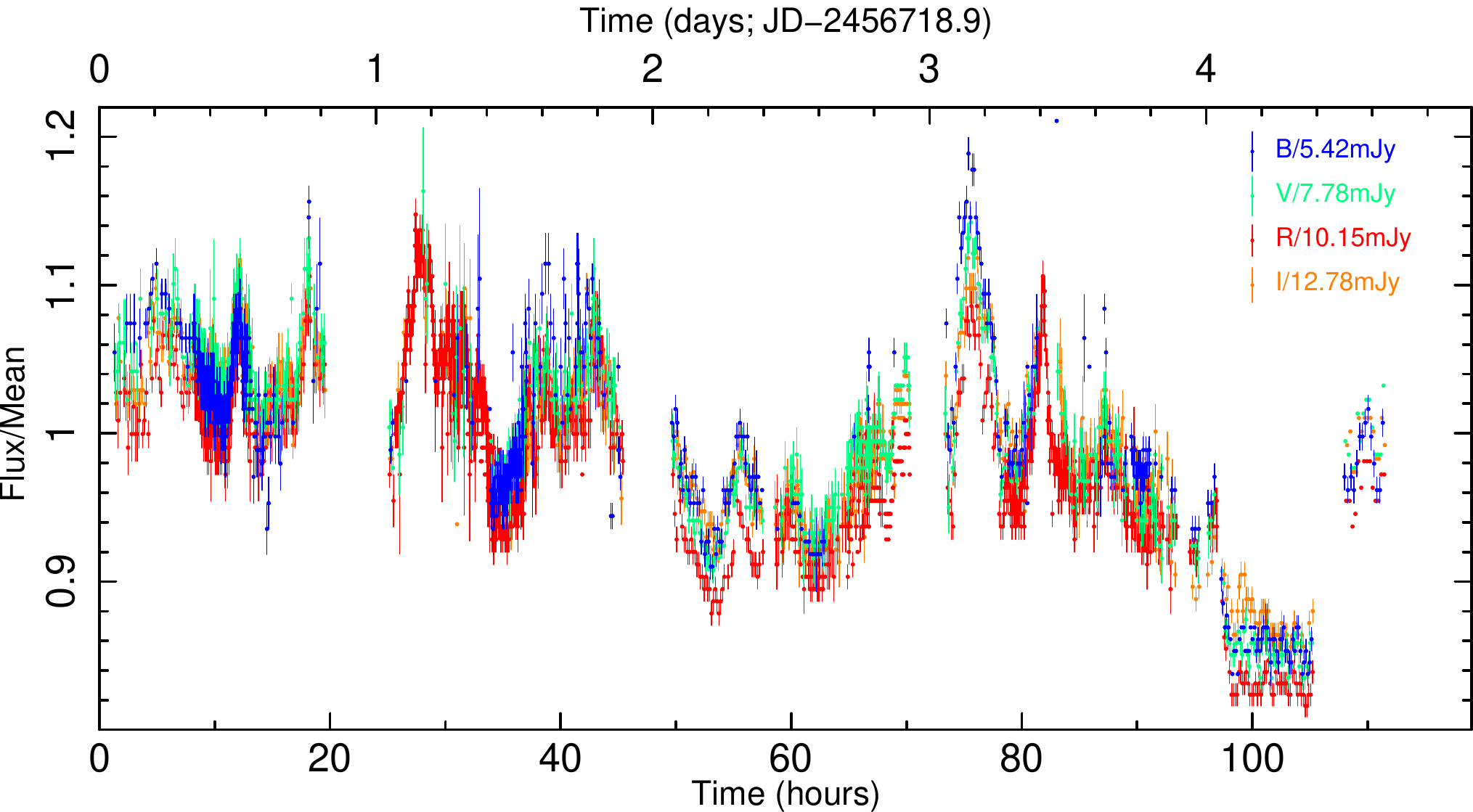}
\caption{Mean-normalized photometric light curves of S5~0716+714 in BVRI filters (see the upper panels in Figure\,\ref{rawlightcurve}) facilitating a visual comparison of variability across the four bands.}
\label{BVRI}
\end{figure*}

\subsection{Photometric data analysis}

The full-campaign mean-normalized light curves in BVRI filters are presented in Figure\,\ref{BVRI}. The source brightness in magnitudes was converted into the flux in mJy units by using the zero points for UBVRI-JHK Cousins-Glass-Johnsons system given in Table A2 of \citet{bessel98}, and to calculate the optical spectra the fluxes were interstellar-extinction corrected using the extinction magnitudes for various filters listed in the NED\footnote{www.ned.ipac.caltech.edu}. As shown in the figure, the photometric data spanned about 112 hours from the start of the campaign, with some interruptions at six locations in time resulting from bad weather conditions and/or a change in active observatories. The corresponding six interruptions were $5.64, 4.33, 1.13, 3.12$ and $2.96$\,h-long, making the net observation exposure 92.83\,h. For about 6 hours, during 99.03 -- 105.22 h, the source suddenly
exhibited a strongly reduced level of flux variability, resulting in a ``plateau" in all four bands' light curves, as seen in Figure\,\ref{BVRI}.  The resulting variability duty cycle, excluding this ``plateau" period, is thus $\sim93\%$.  A detailed discussion on this reduced activity will be presented in \S\, \ref{plateau}.

\begin{table}[bh!]
\caption{Variability amplitudes of S5~0716+714 during the 2014 WEBT campaign.}
\label{table:VA}
\centering

\begin{tabular}{c c c c c }
	&	&Photometric Data &  & \\
\end{tabular}
\begin{tabular}{c c c c c}
\hline\hline
	Filter	&	Number of obs.&Mean Mag. & VA (mag)& $F_{var}$ (\%)\\
\hline
 	B	&	561	&	14.78	&0.38&6.54 $\pm$0.07\\
	V	&	 776	&	14.26	&0.35&5.74 $\pm$0.06\\
	R	&	1921	&        13.79   &0.36&5.79 $\pm$0.03\\
	I	&	723	&	13.28	&0.28&5.28 $\pm$0.05\\
\hline 
\end{tabular} 

\begin{tabular}{c c c } \\
	& Polarimetric Data: Epoch\,I (25--49\,h) & \\ 
\end{tabular}
\begin{tabular}{c c c} 
\hline\hline  
	Obs.	&	Range& $F_{var}$ (\%) \\ 
\hline
 	Flux (mag)	&	13.64 -- 13.86	&4.34 $\pm$ 0.07	\\
	PD (\%)	&	 1.32 -- 10.45	&25.70 $\pm$ 1.00	 \\
	PA (deg.)	&	40.15 -- 75.02	& 10.06 $\pm$ 0.55    \\
\hline
\end{tabular}

\begin{tabular}{c c c} \\
	& Polarimetric Data: Epoch\,II (79--97\,h) & \\ 
\end{tabular}
\begin{tabular}{c c c} 
\hline\hline
	Obs.	&	Range& $F_{var}$ (\%) \\ 
\hline                  
 	Flux (mag)	&13.66 -- 13.88 	&3.90 $\pm$ 0.05\\
	PD (\%)	&3.45 -- 12.36	&27.90 $\pm$ 0.30 \\
	PA (deg.)	& 13.59 -- 42.25 & 22.58 $\pm$ 0.37\\
\hline
\end{tabular}

\end{table}

Of the four filters analyzed, the data in the B filter have the largest scatter and the least number of data points, whereas the data in filter R have the least scatter and the largest number of data points. The amplitude of the peak-to-peak variations was estimated by using the relation given in \citet{Heidt96},
\begin{equation}
{\rm VA} = \sqrt{(A_{max}-A_{min})^2-2\sigma ^{2}} \, , 
\end{equation}
where $A_{max}$, $A_{min}$, and $\sigma$ are the maximum, minimum, and standard deviation of the light curve, respectively. However, the estimation of this amplitude considers only the two extreme flux measurements, and hence may not represent the overall variability during the campaign. Fractional variability $F_{var}$, on the other hand, includes all the observations and hence provides a better index for the overall variability of the source \citep[see][]{vau03,Edelson02}. Both of these parameters are listed in Table\,\ref{table:VA} for BVRI filters. 

\subsubsection{Characteristic variability timescales}

Study of characteristic variability time scales of blazar light curves proves to be one of the most important tools that can be used to constrain sizes and geometrical structures of blazar emission zones. Small-amplitude flux changes with typical durations of about a few hours, are very likely to originate in the closest vicinities of supermassive black holes launching the jets, and as such may be shaped by a combination of accretion disk instabilities, MHD waves propagating within the outflow, and/or particle acceleration and radiative cooling timescales at the jet base, etc. \citep[see, e.g.,][]{Ulrich97}. A proper characterization of such time scales, along with the search for quasi-periodic oscillations (QPOs), was in fact one of the key motivations to conduct the 2014 WEBT campaign targeting S5~0716+714.

\begin{figure}[t!]
\centering
\includegraphics[width=\columnwidth]{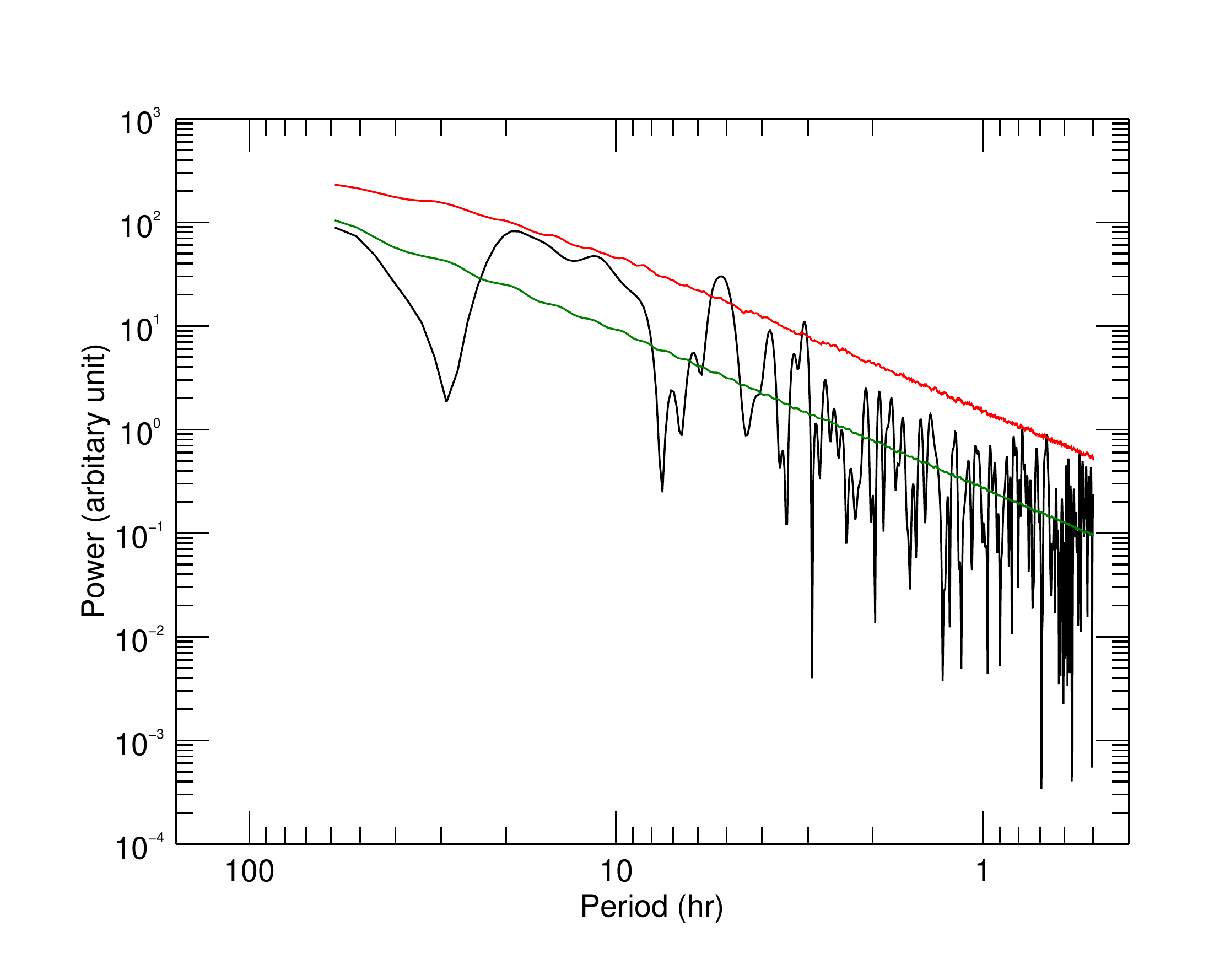}
\caption{The LS periodogram of S5~0716+714 (for the duration of the 2014 WEBT campaign) in R filter (black curve), along with the mean periodogram (green curve) and the 99\% significance curve (red curve) from the MC simulation.}
\label{fig:LSP}
\end{figure}

We carried out frequency-domain analysis of the source light curves, as prescribed in \citet{Lomb76} and \citet{Scargle76}, and searched for significant peaks corresponding to possible QPOs. Lomb-Scargle (LS) periodogram is considered to be a powerful method allowing to detect and to test the significance of a periodic signal in unevenly sampled and noisy time series. The method, although similar to the ordinary discrete periodogram in many respects, relies on a different approach to spectral analysis, as it estimates the spectral power by the least-square fitting of the data with a model function of the type $y(t)=A\, \sin\omega t+B\, \cos \omega t$. The upper panel in Figure\,\ref{fig:LSP} presents the resulting LS periodogram for S5~0716+714 in the best-sampled R filter. As revealed by the plot, oscillations with periods of $\simeq 3$\,h and $\simeq 5$\,h \emph{could possibly} be significant enough to indicate the presence of QPOs in the source light curve. 

It is important to realize that, however, any analysis of real time series, including the LS periodogram, may be subjected to ``spectral leakage'' and ``aliasing'', due to the fact that the analyzed light curve is finite in time, and due to intervals between two successive measurements, in particular in the case of a frequency-dependent (red) noise type of a source variability; similarly, all the monitoring breaks and gaps, unavoidable in any astronomical time series, may distort further the analysis results by introducing spurious peaks in the periodogram \citep[see in this context][]{Press78}. Therefore, the presence of QPOs in the analyzed light curve should be investigated rigorously. Hence, to estimate the true significance of the peaks present in the LS periodogram, we conducted a significance test using a large number of simulated light curves based on a modeled power-spectral density (PSD) function, following the method by \citet{TK95}. The method relies on randomizing both the phase and amplitude of the Fourier transform coefficients, in order to account for the observed statistical behavior of the periodogram. 

\begin{figure}[t!]
\centering
\includegraphics[width=\columnwidth]{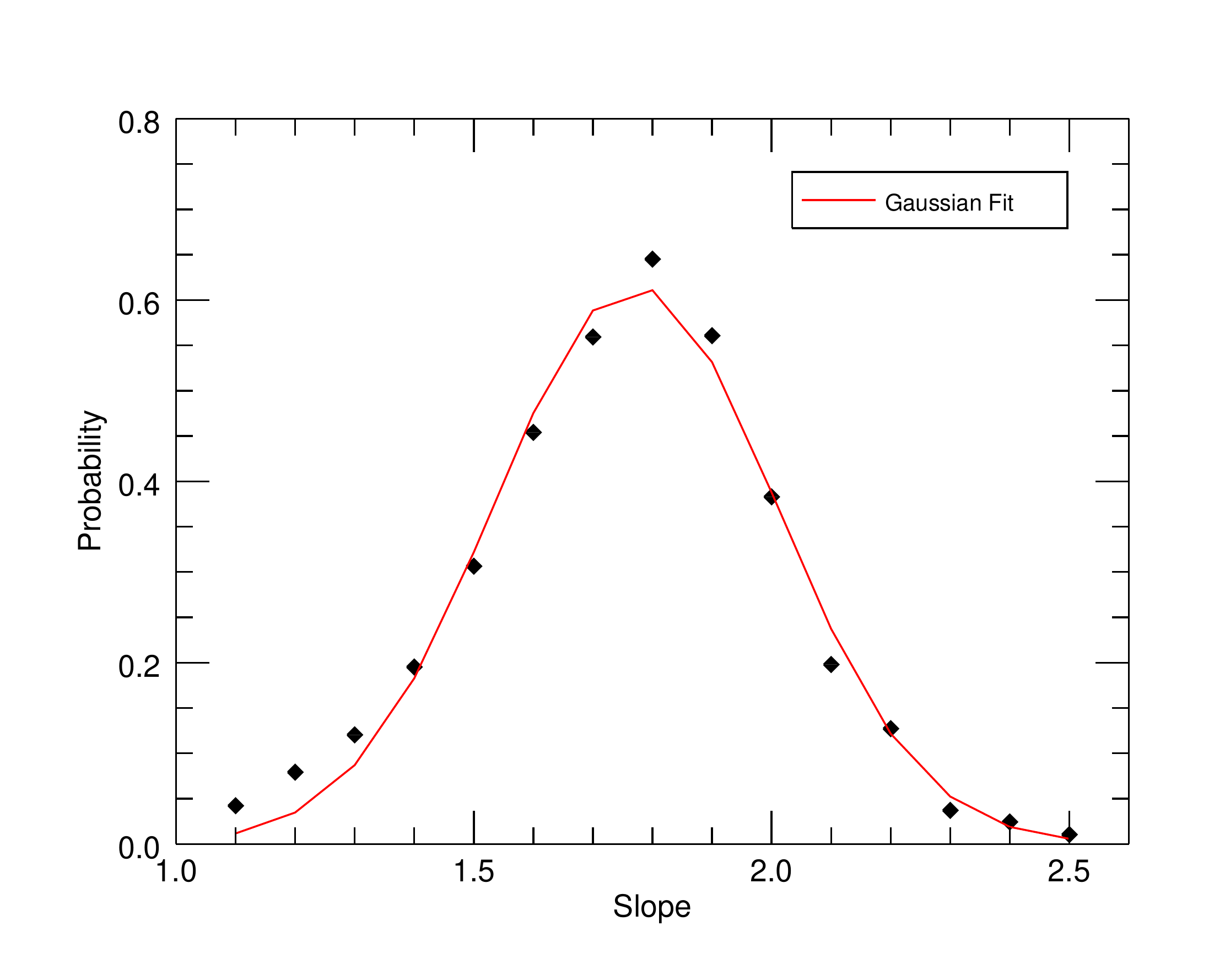}
\includegraphics[width=\columnwidth]{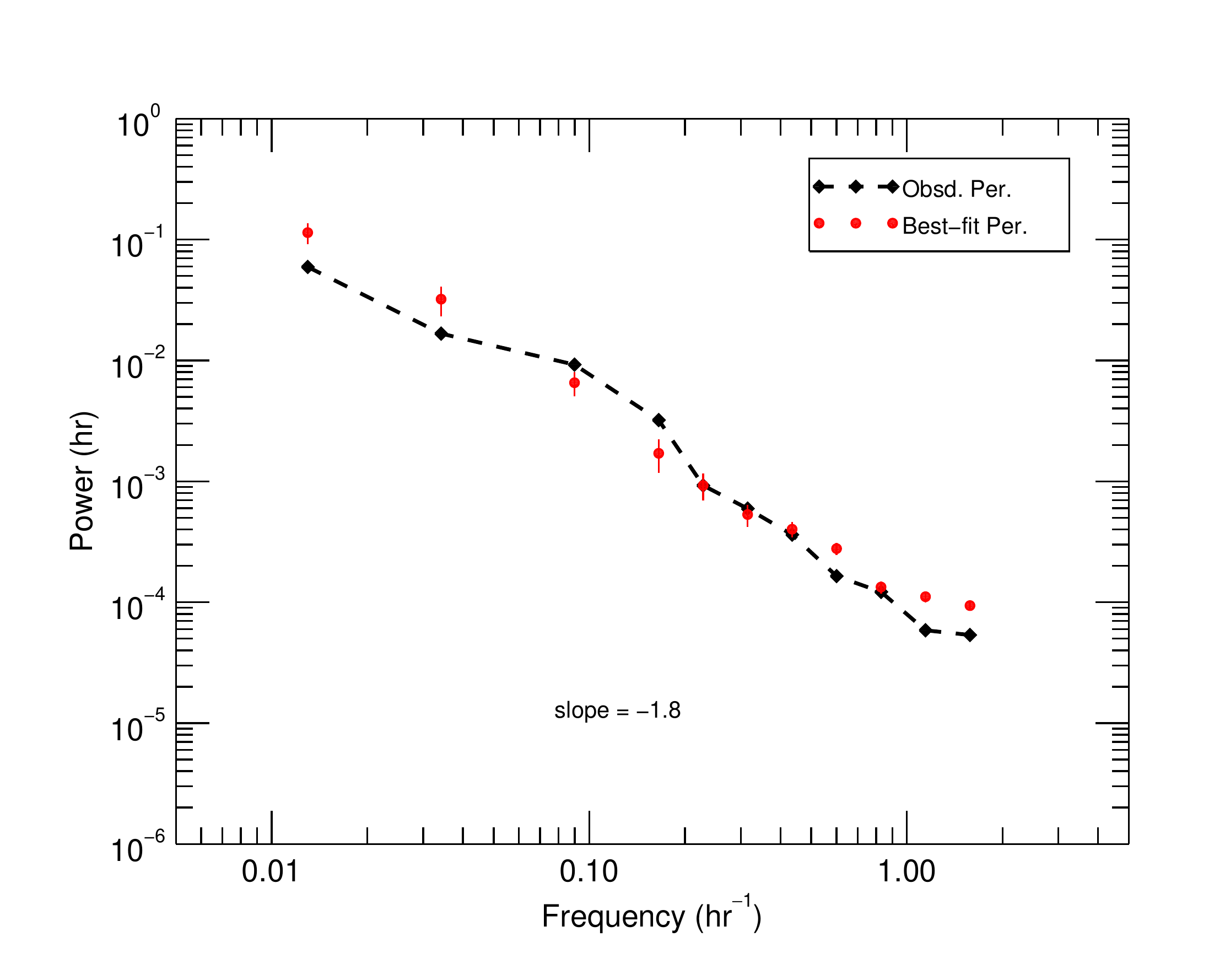}
\caption{{\it Upper panel}: Probability distribution of the PSD slopes for S5~0716+714 (for the duration of the 2014 WEBT campaign) in R filter (black symbols); the red solid line denotes the corresponding Gaussian fit. {\it Lower panel}: Binned periodogram of S5~0716+71 in R band (black symbols connected by a dotted curve), along with the average of $1,000$ binned periodograms simulated using the best-fit model slope of $\beta = 1.8$; the errors give standard deviation of the simulated periodograms from the average.}
\label{fig:fit_slope}
\end{figure}

First, we estimated the parameters of the PSD, assuming a power-law model that best represents the observed periodogram, according to the power-response method (PSRESP) described in \citet{Uttley02}, which has been widely used in the analyses of AGN variability in general \citep[e.g.,][]{Chatterjee08,max14,Chen16}. Here we briefly summarize the method as follows:
\begin{itemize} 
\item[i)] For a given time series $f(t_{j})$ sampled at times $t_{j}$ with $j = 1, 2,.., N$, the discrete Fourier power at an angular frequency $\omega$ was estimated using the expression
\begin{equation}
P\!\left(\nu \right)=\frac{2\,T}{\left(N \bar{f} \right)^{2}} \, \left | \sum_{j=1}^{N}f\!\left( t_{i} \right ) \, e^{-i2 \pi \nu t_{j}} \right |^{2} \, ,
\end{equation}
where $T$ and $\bar{f}$ represent the total duration of the series, and the mean flux of the source, respectively; the periodogram was binned using suitable frequency bins, so as to reduce the scatter in the periodogram for a model fitting.
\item[ii)] Based on an arbitrary single power-law model $P(f) = N_{0} \times f^{-\beta }+C$ with the added Poisson noise, $1,000$ source light curves were simulated with the given sampling of the data $f(t_{j})$; subsequently, for each simulated light curve binned Discrete Fourier Transform (DFT) periodogram was estimated using the same binning as for the data.
\item[iii)] For each of the simulated light curve, a $\chi^{2}$-like quantity (not the same as the conventional $\chi^{2}$)  was calculated using the expression
\begin{equation}
\chi_{i}^{2}=\sum_{\nu _{\rm min}}^{\nu _{\rm max}}\frac{\left [ \overline{P_{\rm sim}\left ( \nu  \right )} -P_{i}\left ( \nu  \right )\right ]^{2}}{\Delta \overline{P_{\rm sim}\left ( \nu  \right )}^{2}},
\label{chi}
\end{equation}
\noindent where $ \overline{P_{\rm sim}\left ( \nu  \right )}$ and $\Delta \overline{P_{\rm sim}\left ( \nu  \right )}$ stand for the mean periodogram and the standard deviation of the $1,000$ periodograms of the simulated light curves; a similar quantity for the observed periodogram, $\chi_{\rm obs}^{2}$, was also evaluated by replacing $P_{i}$ with $P_{\rm obs}$.
\item[iv)] Step iii) was repeated for 15 various slopes of the power-law model.
\item[v)] The goodness of fit between the mean simulated periodogram and the observed periodogram was estimated by comparing $\chi^{2}_{\rm obs}$ with $\chi^{2}_{i}$s; in particular, the ratio of the number of $\chi^{2}_{i}$s greater than $\chi^{2}_{obs}$ to the total number of $\chi^{2}_{i}$s in all ($15 \times1,000$) simulations defined the probability used to quantify the goodness of the fit for a given model. In a situation where the fit statistics is not well-understood, such a method involving the use of simulated data for the estimation of goodness of fit is well understood and discussed in \citet[][section 15.6]{Press92}
\end{itemize} 
The resulting probability distribution of the PSD slopes for S5~0716+714 (for the duration of the 2014 WEBT campaign) in R filter is presented in the upper panel of Figure\,\ref{fig:fit_slope}. The best-fit slope (with the highest probability of $0.64$) was found to be $\beta = 1.8\pm0.3$, where the half-width at half maximum (HWHM) for the Gaussian fit of the slope distribution was associated with the uncertainty in the slope estimate. During the analysis, the slope index, being the primary parameter of interest, was the only parameter varied; the other parameters $N_{0}$ and $C$ were fixed to $0.97\ h^{-1}$  and $10^{-4}\ h$, respectively. The lower panel in Figure\,\ref{fig:fit_slope} shows the binned mean simulated periodogram with the slope index $1.8$, and the binned observed periodogram of the source.

Next, with the given best-fit power-law model of the PSD, we simulated $10,000$ light curves which were then re-sampled to match the sampling of the observed light curve of the source. Subsequently, the distribution of LS periodograms of the simulated light curves were used to estimate the significance of the QPO-like features. The average of the simulated light curves is shown in the upper panel of Figure\,\ref{fig:LSP} (green curve), along with the 99\% confidence level curve (red curve). The analysis indicates that the power around the periods of $3.05\pm0.14$\,h and $5.17\pm0.52$\,h is significant at the level of $99.68$ and $99.91\%$, respectively. The uncertainties (Gaussian fit HWHMs) associated with the periods of the QPO-like features were estimated by subtracting the simulated mean power level from the observed power.

On the other hand, one should note that the $99 \%$ confidence level derived above denotes the ``single-trial'' confidence bound, i.e. the probability that a periodogram point will exceed this height under the assumption that the null hypothesis model (here: pure red-noise PSD with a power-law slope of 1.8) is correct.  We now attempt to estimate the ``global'' $99 \%$ confidence bound, accounting for the fact that we searched over a large number of frequencies. However, the lack of complete independence of neighboring frequencies
in the LS periodogram means that the confidence bounds given by \citet[section 4 therein]{Vaughan05} \emph{cannot} be used at face value, since they were derived for the limit of strictly even sampling.

\begin{figure}[t!]
\centering
\includegraphics[width=\columnwidth]{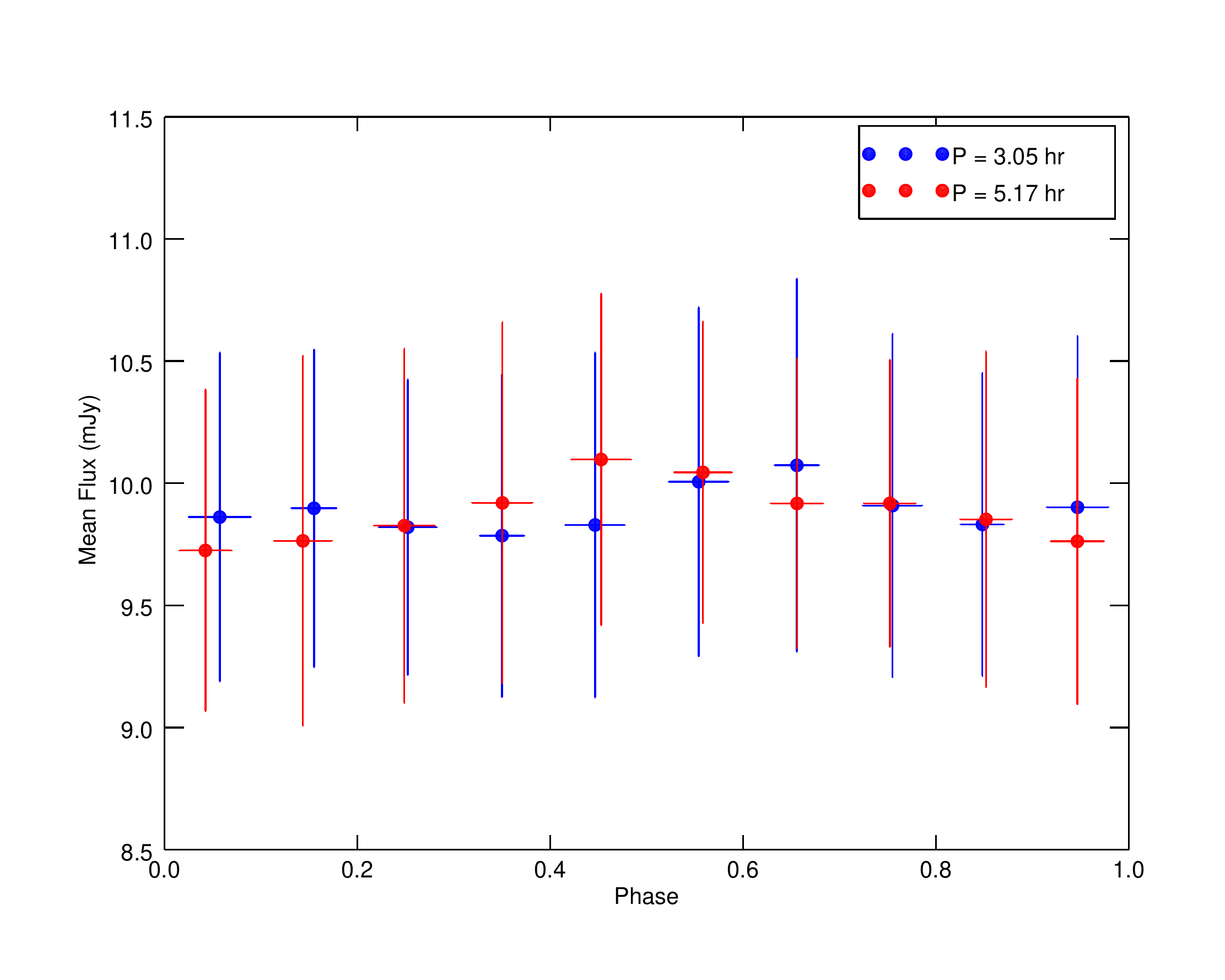} 
\caption{The folded light curve of S5~0716+714 with the periods of $3.05$\,h (blue) and $5.17$\,h (red).}
\label{Folded}
\end{figure}

We find empirically at selected frequencies that the distribution of our LS periodogram points usually follows a rough exponential distribution, but the $99 \%$ single-trial confidence bound derived from the simulations indicate a typically $\sim30\%$ larger dispersion compared to the distribution for the case of even sampling ($\chi^2_2$ distribution, i.e., an exponential probability distribution with variance of 4). Defining $z$ to be the ratio of a periodogram point to the true mean PSD at any given frequency, our simulations indicate that the single-trial $99 \%$ confidence bounds typically correspond to values of $z \sim 4-8$. We now make the simplifying assumption that $z=6$ represents the $99\%$ single-trial probability across all frequencies of interest (compared to $z = 4.6$ for the evenly-sampled case). The $99\%$ global confidence bounds can thus be estimated (following $\S$3 of  \citealt{Frescura08}, and paralleling Equation 16 of \citealt{Vaughan05}) as $2z \sim -2.6 \ln (0.01/n')$, where $n'$ denotes the number of independent frequencies. Using the empirical formula of \citet{Horne86} we obtain $n' > 1000$, but this value seems overestimated \citep[see][]{Frescura08}; instead, we take $n'$ to lie in the approximate range 200--800. This range yields a 99$\%$ confidence bound of $z$ approximately 12.8--14.7.

The ``candidate features'' in the LS periodogram at 3\,h and 5\,h correspond to approximately $z = 8$ and $z= 9$, and the global confidences of approximately $58 \%$ and $81 \%$, for $n'=200$, respectively, so we cannot conclude that these features represent significant deviations from the null hypothesis model. This is supported further by the data folding analysis, the results of which are presented in Figure\,\ref{Folded}, which does not reveal any significant pulse profiles corresponding to the two periods analyzed. Hence, If there does exist a characteristic timescale, it could simply lie outside the range searched in this paper. Alternatively, the dominant variability processes in S5~0716+714 over timescales of tens of minutes to a few days are scale-invariant.

\begin{figure}[t!]
\centering
\includegraphics[width=\columnwidth]{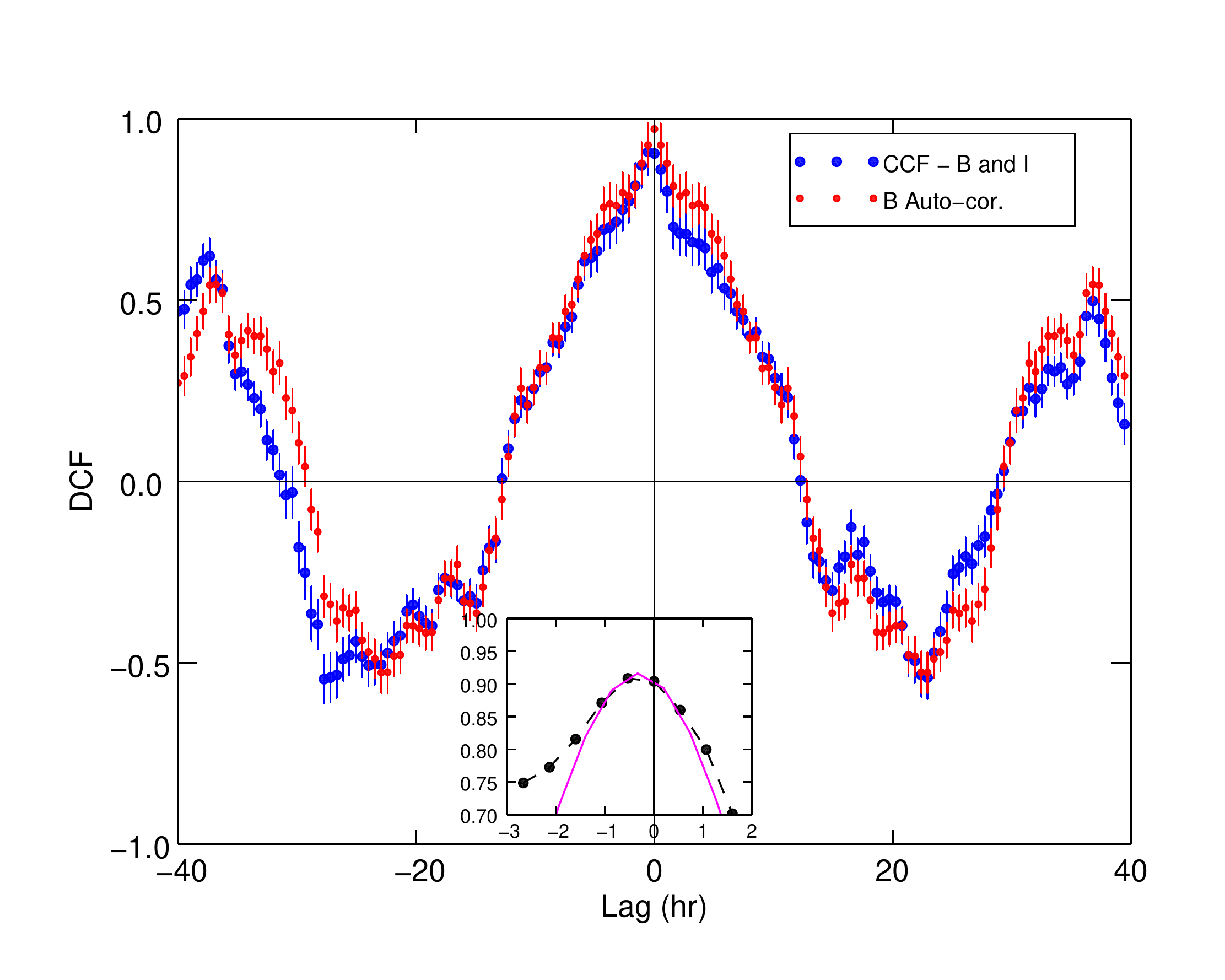} 
\caption{The discrete cross-correlation function (DCF) for S5~0716+714 between B and I fluxes (blue symbols), along with the auto-correlation function (ACF) for the B band flux (red symbols).The inlay plot zooms into the DCF centered around zero lag (black points) and the Gaussian fit (magenta curve). A negative lag here indicates that variations in B-band lead those in I-band.}
\label{CCF}
\end{figure}

\subsubsection{Correlated flux variability}

Cross-correlation analysis between different filters offers an important clue about a structure of the blazar emission region, and the main radiative processes involved. If a statistical significance of any lag between the flux variation in different bands can be established, such lags could for example imply a spatial separation between distinct emission zones dominating radiative output of the source at different frequencies. The discrete correlation function (DCF) discussed in \citet{EK88}, is one of the most extensively used methods to investigate the cross-correlation between two time series with uneven spacing. However, in this method the maximum and minimum DCF, not being standardized, the normalization given in \citet{Welsh99} was applied to limit the DCF values between $\ -1$ and $\ +1$ as in standard correlation function.  We calculated the normalized DCF between B and I light curves, which are the bands with the largest wavelength separation in the 2014 WEBT campaign (excluding the JHK ones that span only a few hours).  The DCF between B and I light curves and the auto-correlation function (ACF) for B light curve for total lag about a half of the total time span of observations are shown in Figure\,\ref{CCF}.  In the figure, the striking resemblance between DCF and ACF suggests that the light curves are highly correlated over the period of time. However, the inlay plot reveals that there could be  a marginal lead of the B-band emission over the I-emission by $\sim 0.6\pm 0.11$\,h (the error estimated by HWHM of the Gaussian fit). 

\begin{figure}[t!]
\centering
\includegraphics[width=\columnwidth]{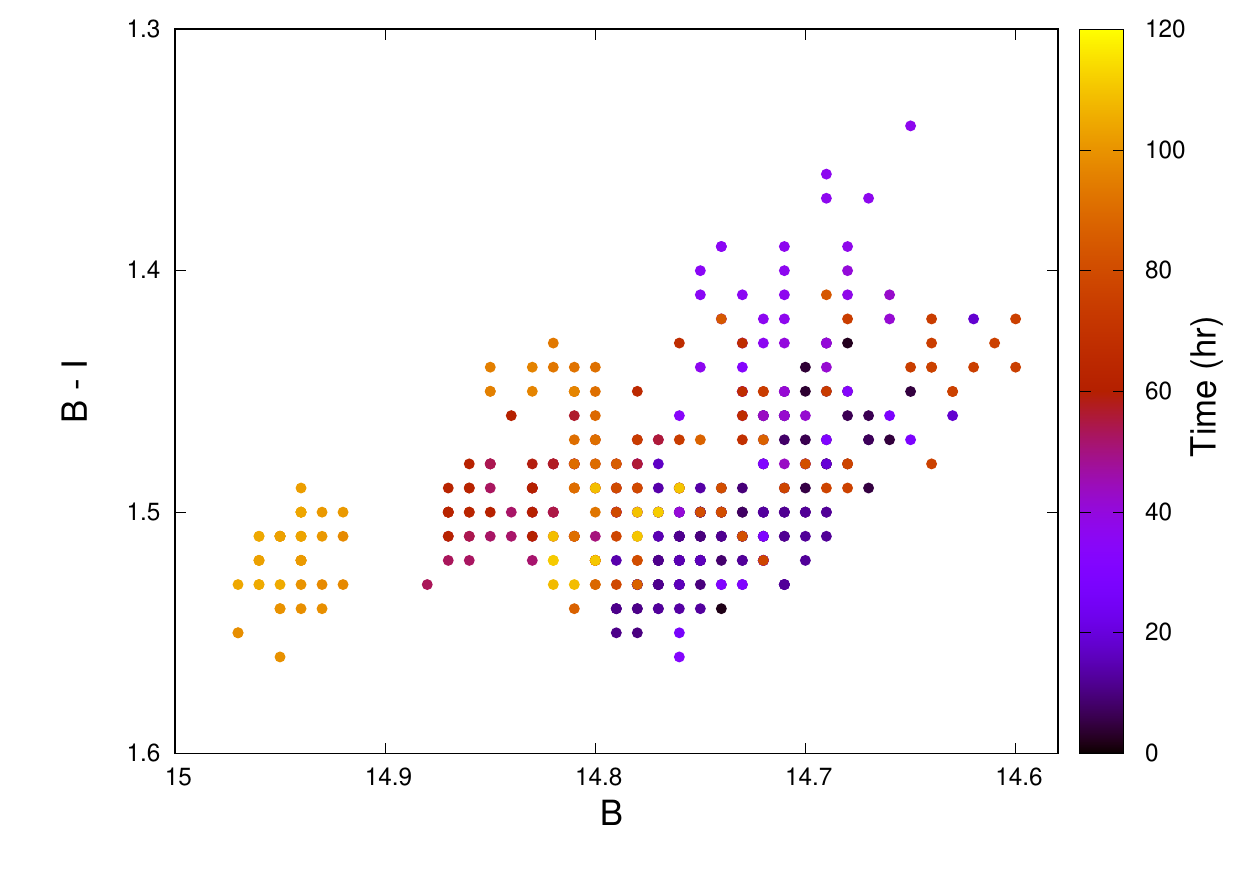} 
\includegraphics[width=\columnwidth]{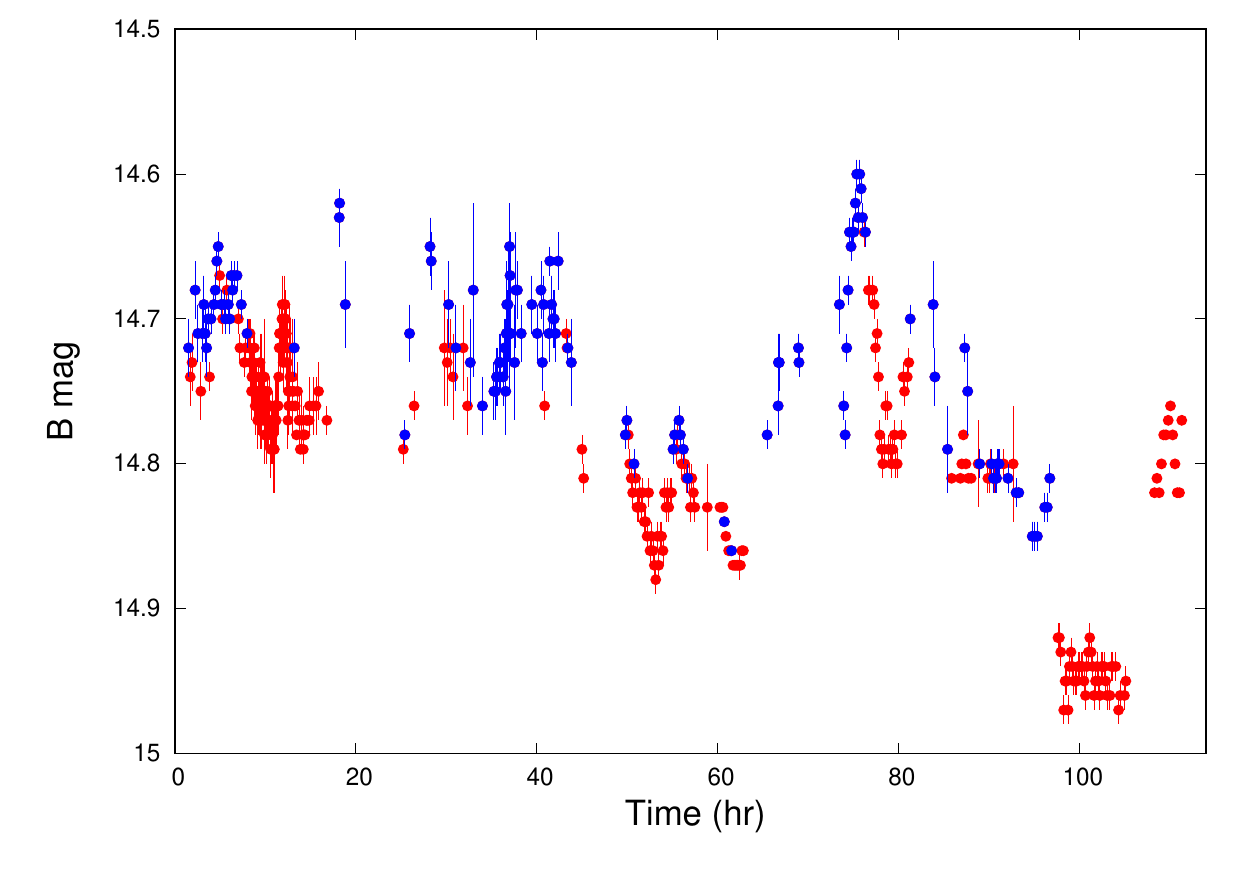} 
\caption{{\it Upper panel}: Color B-I vs. B magnitude diagram for S5~0716+714 during the 2014 WEBT campaign; the plot is color-coded so that the observing time runs from blue to yellow; the errors in color and magnitude are not shown for clarity. {\it Lower panel}: The corresponding B-band light curve of the source, for which blue symbols correspond to flat spectra, defined by the lower 30 percentile $B - I$ color value, $1.48$, and red symbols to steep spectra, i.e. larger values of $B - I$.}
\label{colorMag}
\end{figure}

\subsubsection{Color variability}

During the campaign, the source exhibited not only flux variability, but also showed some (relatively moderate) variation in color between B and I bands ($\sim 1.35$ mag), the widest spectral window in the 2014 WEBT data. The  apparent correlation between the B flux and the $B - I$ color is shown in the upper panel of Figure\,\ref{colorMag}. The figure is color-coded, so that the observing time runs from blue to yellow. The bottom panel of the figure presents the B-band light curve of the source, for which blue symbols correspond to flat spectra, defined by the lower 30 percentile B-I color value, $1.48$, and red symbols to steep spectra, i.e. larger values of B-I. As shown, flux maxima appear bluer than flux minima for the analyzed light curve, equivalently to the ``bluer-when-brighter'' trend claimed for S5~0716+714 already in the past \citep[e.g.,][]{Ghisellini97,Dai13}, and found in other BL Lacs as well \citep[e.g.,][and references therein]{Ikejiri11,Wierzcholska15}.

In general, bluer-when-brighter behavior is indicative of a connection between between the observed flux enhancement and the episodes of an intensified particle acceleration within the emission site. Purely geometrical in nature changes in the flow beaming pattern, which are expected to lead to rather achromatic flux variability, could not account for the observed trend. Alternatively, spectral flattening witnessed during the elevated flux levels could be explained assuming an underlying steady electron energy spectrum of a curved/concave shape, superimposed on a strongly fluctuating (i.e., occasionally compressed, or amplified) magnetic field; local enhancements in the jet comoving magnetic field intensity $B'$ would then lead to an increased synchrotron emissivity at a given observed frequency, produced by the electrons with correspondingly lower energies $E_e \propto 1/\sqrt{B'}$, and therefore flatter spectrum.

\begin{figure}[t!]
\centering
\includegraphics[width=\columnwidth]{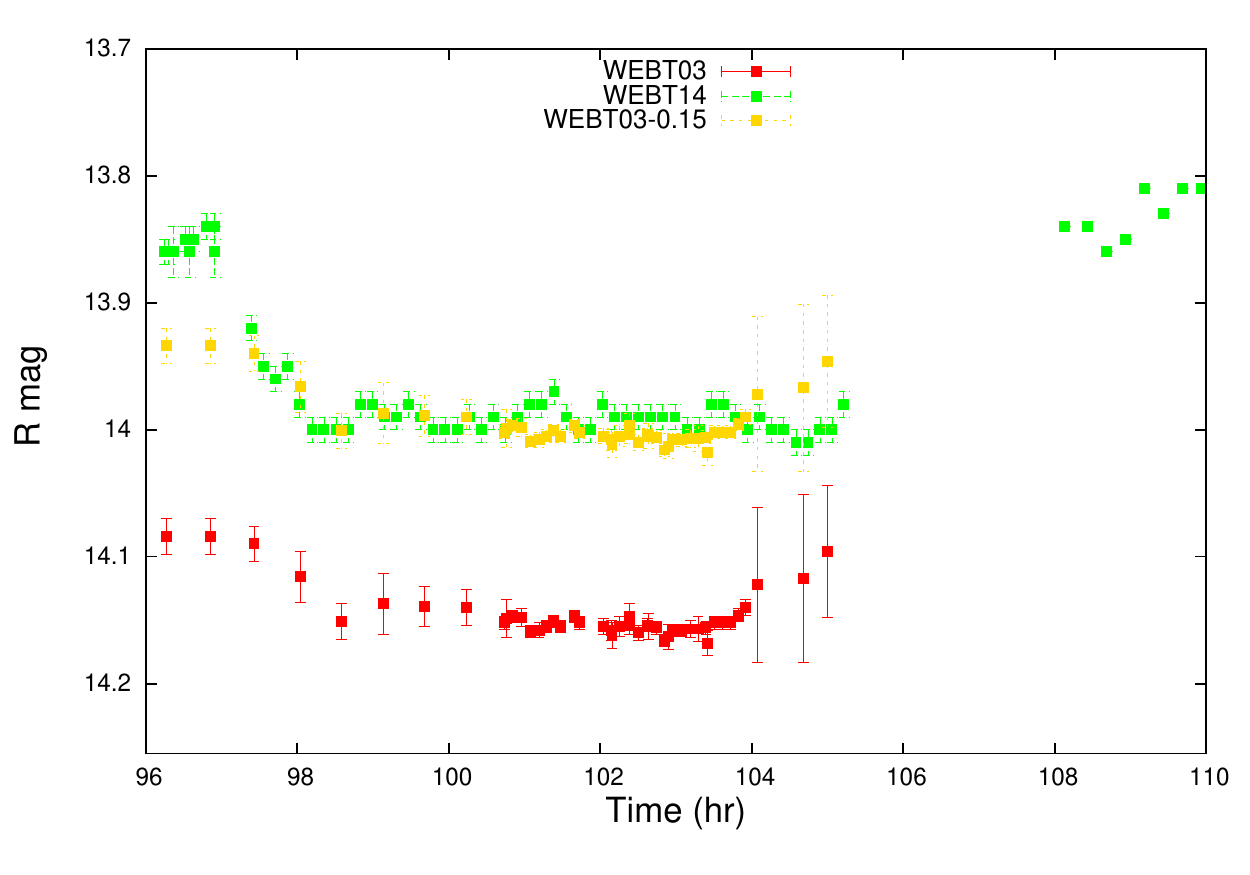} 
\caption{2014 WEBT light curve of S5~0716+714 in R band during the plateau phase (green symbols), compared with the analogous event detected during the 2003 WEBT campaign \citep[JD 2452956.38325 --  2452956.74681; see][red symbols]{Ostorero06}, and the same segment of the 2003 light curve just shifted vertically by $-0.15$\,mag (yellow symbols).}
\label{webt06}
\end{figure}

\subsubsection{The plateau} \label{plateau}

It is interesting to note in Figure\,\ref{BVRI} that, even though the light curves in all four filters undergo pronounced variations throughout the entire campaign period, as expected in the case of S5~0716+714 famous for its very high flaring duty cycle, at around the 97th hour from the beginning of the 2014 WEBT observation the source suddenly dimmed at all the frequencies by a few tenths of magnitudes, and remained at a constant (low) flux level for about $6$ hours. In R filter, the flux dropped in particular by 0.15\,mag down to $\sim 14.0$\,mag.  Values of $F_{\rm var}$ during the plateau period spanned  1.20--1.33 $\pm$ 0.14--0.16 $\%$ across the four bands;   locally (over $\sim6$ hr timescales), $F_{\rm var}$ was typically  $\sim2-6\%$ at most other periods in the light curves.

\begin{figure}[t!]
\centering
\includegraphics[width=\columnwidth]{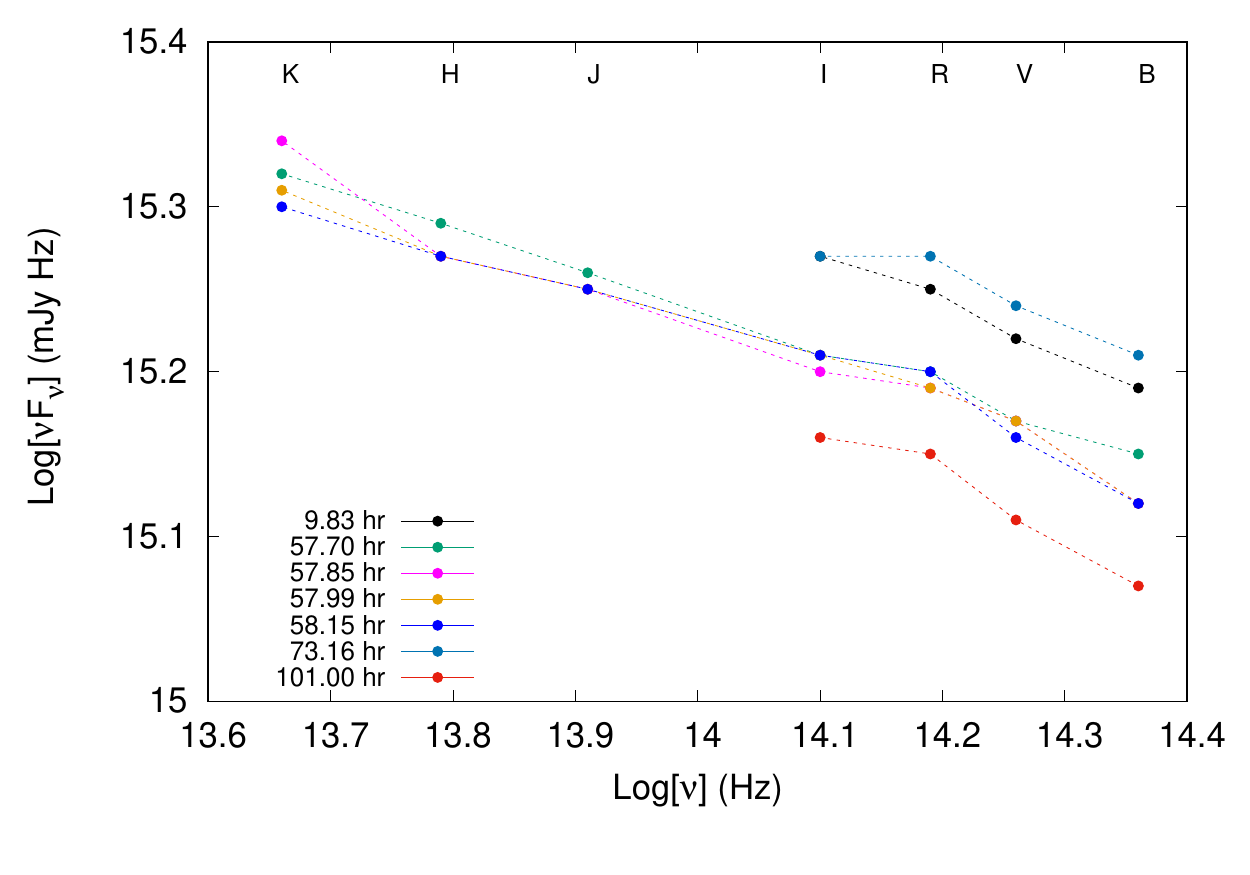}
\caption{Optical spectra of S5~0716+714 during the 2014 WEBT campaign, at different times of the observations, as indicated in the plot. The letters on the plot represent the filters used. The average spectral slope is $\alpha \simeq 1.2$.}
\label{spectra}
\end{figure}

To make sure that this is not an instrumental artifact, we repeated the photometry with the original images several times and checked carefully the data for possible errors. Interestingly, we found a strikingly similar episode of temporary source inactivity in the 2003 WEBT campaign data discussed in \citet{Ostorero06}. The R flux at that time fell by about 0.2\,mag in about $\sim 2$\,h down to 14.15\,mag, and remained constant for about $6$\,h. The corresponding segments of the source light curve from both 2003 \citep{Ostorero06} and 2014 (this paper) WEBT campaigns, are presented in Figure\,\ref{webt06}. Surprisingly, no substantial change in the spectral slope was observed during the plateau phase, as shown in Figure\,\ref{spectra}, indicating that the observed flux during the plateau phase --- a power-law with the spectral index $\gtrsim 1$ --- is still dominated by the jet, and not, for example, by the accretion disk emission.

\subsection{Photo-polarimetric data: Multivariable analysis}
\label{sec:pol}

Apart from the photometric data, the campaign resulted also in the polarimetric data sampled densely in R filter (in addition to a few single measurements in B, V and I filters; see Figure\,\ref{rawlightcurve}). The two well-covered epochs with such polarimetric data correspond to the time intervals from 25th to 39th, and from 79th to 97th hour from the start of the observation, hereafter referred to as 14\,h-long ``Epoch\,I'' and 18\,h-long ``Epoch\,II'', respectively. A detailed study of correlations between the flux, PD, and PA during these epochs, is presented in the following sub-sections.

\subsubsection{Correlations between flux, PD, and PA}

In order to investigate the correlation between the observed variations in flux, PD, and PA, we carried out the DCF analysis for the photo-polarimetric data in R band collected by the AZT-8, LX-200, Perkins and Kanata telescopes for both Epoch\,I and Epoch\,II. We note that the large error bars that can be seen in the first part of the Kanata polarization data, are due to the ongoing maintenance of the reflector of the telescope. 

\begin{figure}[t!]
\centering
\includegraphics[width=\columnwidth]{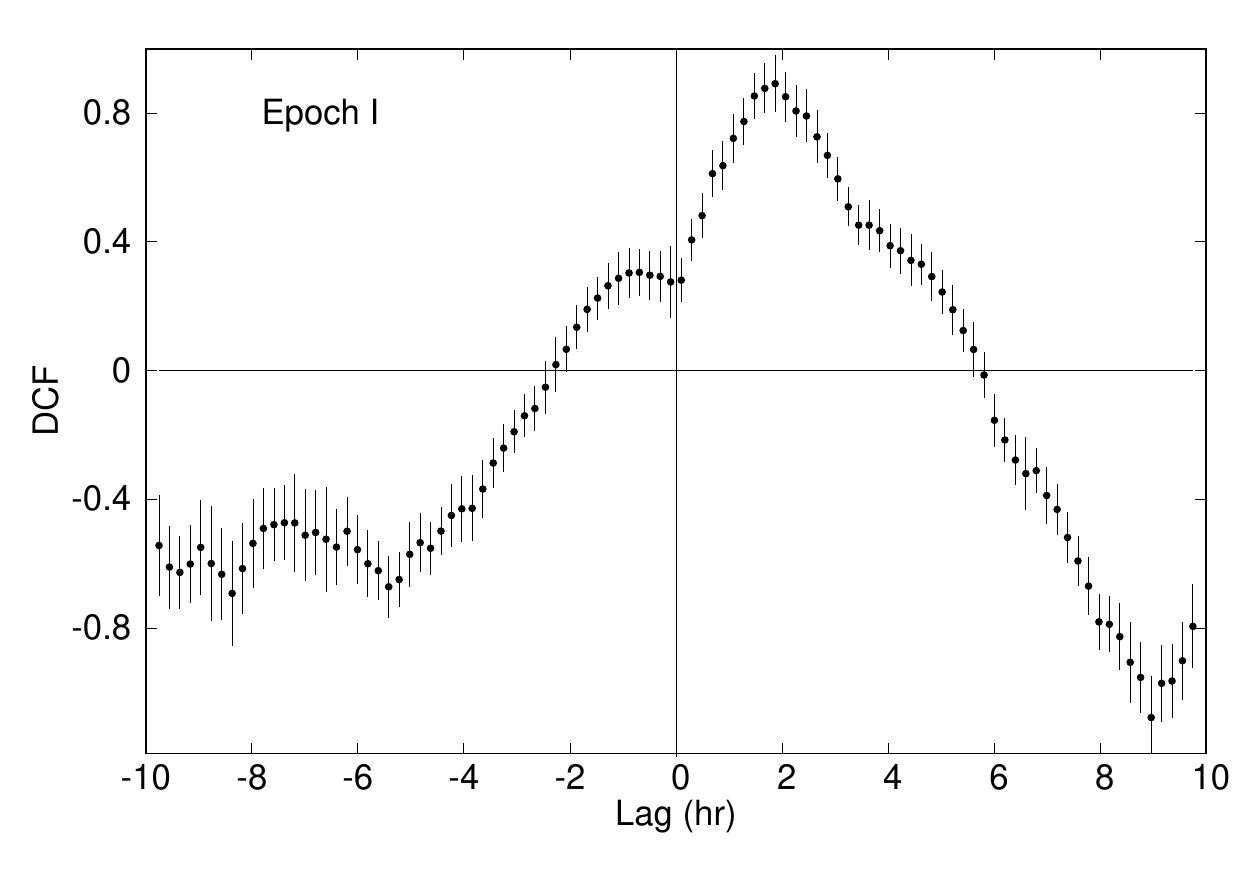} 
\includegraphics[width=\columnwidth]{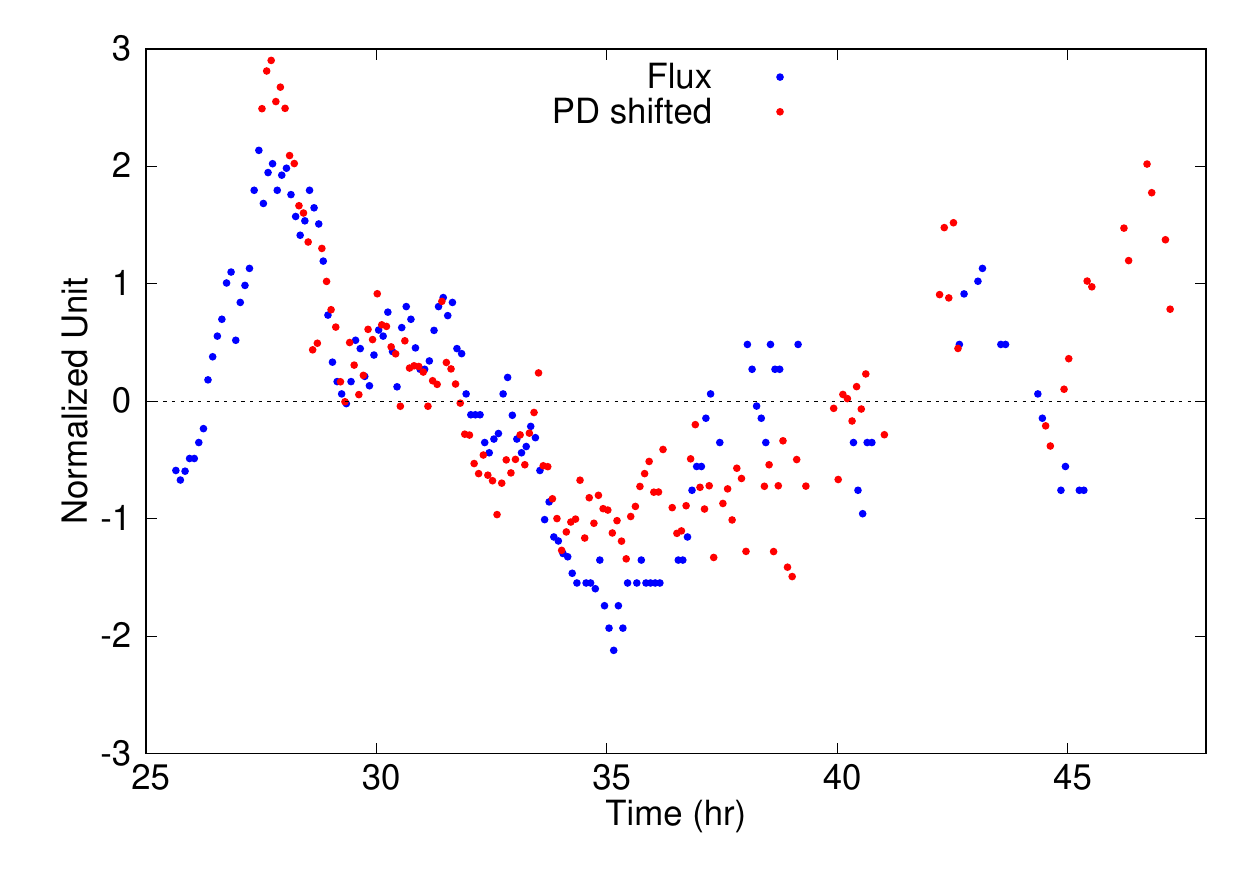}  
\includegraphics[width=\columnwidth]{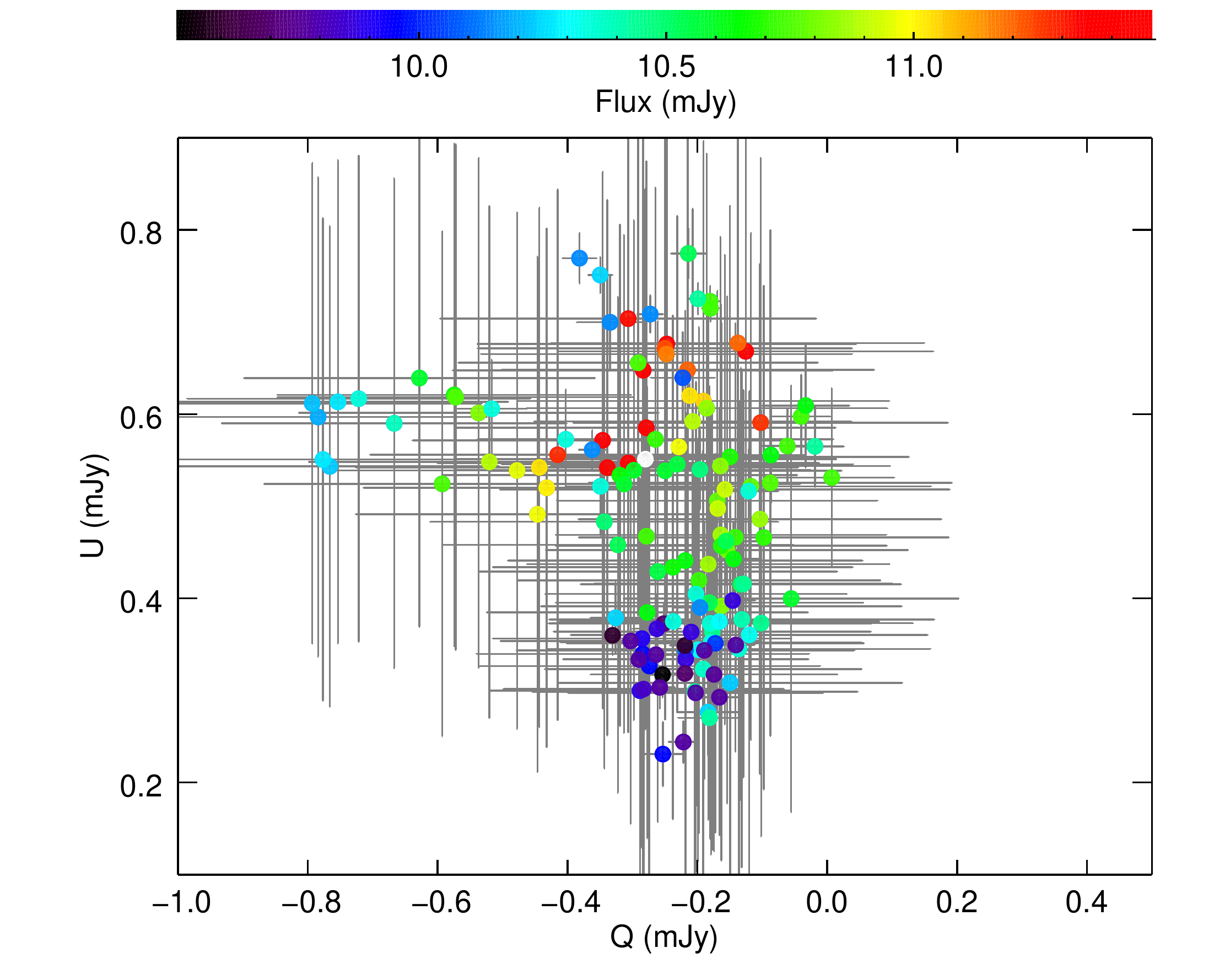}
\caption{{\it Upper panel}: DCF between PD  and R flux during Epoch I. A positive lag indicates PD changes are leading the flux variations. {\it Middle panel}: The corresponding normalized R-band flux light curve (blue symbols), and the PD light curve shifted horizontally by 1.9\,h (red symbols). {\it Lower panel}: The corresponding source evolution on the $Q-U$ Stokes parameters plane. The color scale, from purple to red, indicates the corresponding total flux state from low to high.} 
\label{cor_1}
\end{figure}

\begin{figure}[t!]
\centering
\includegraphics[width=\columnwidth]{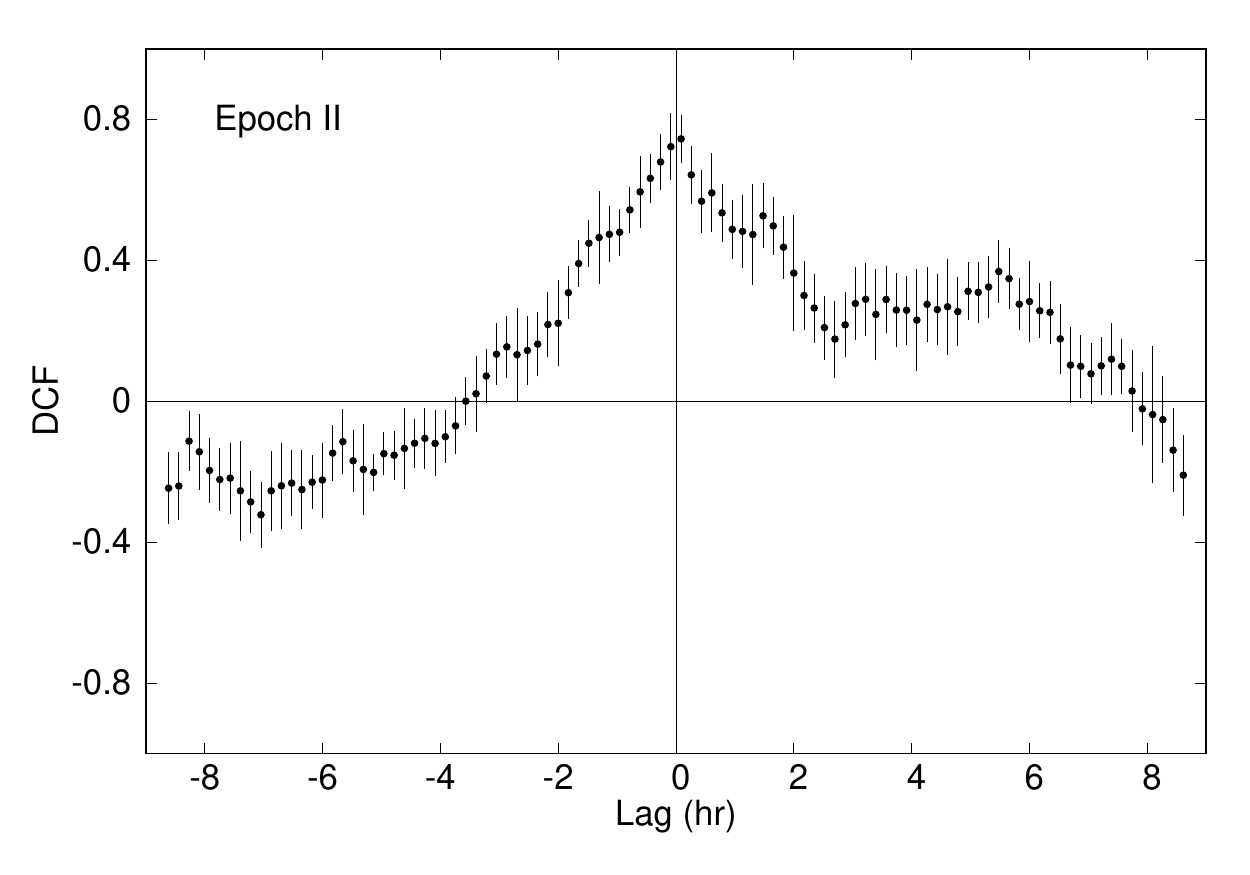}   
\includegraphics[width=\columnwidth]{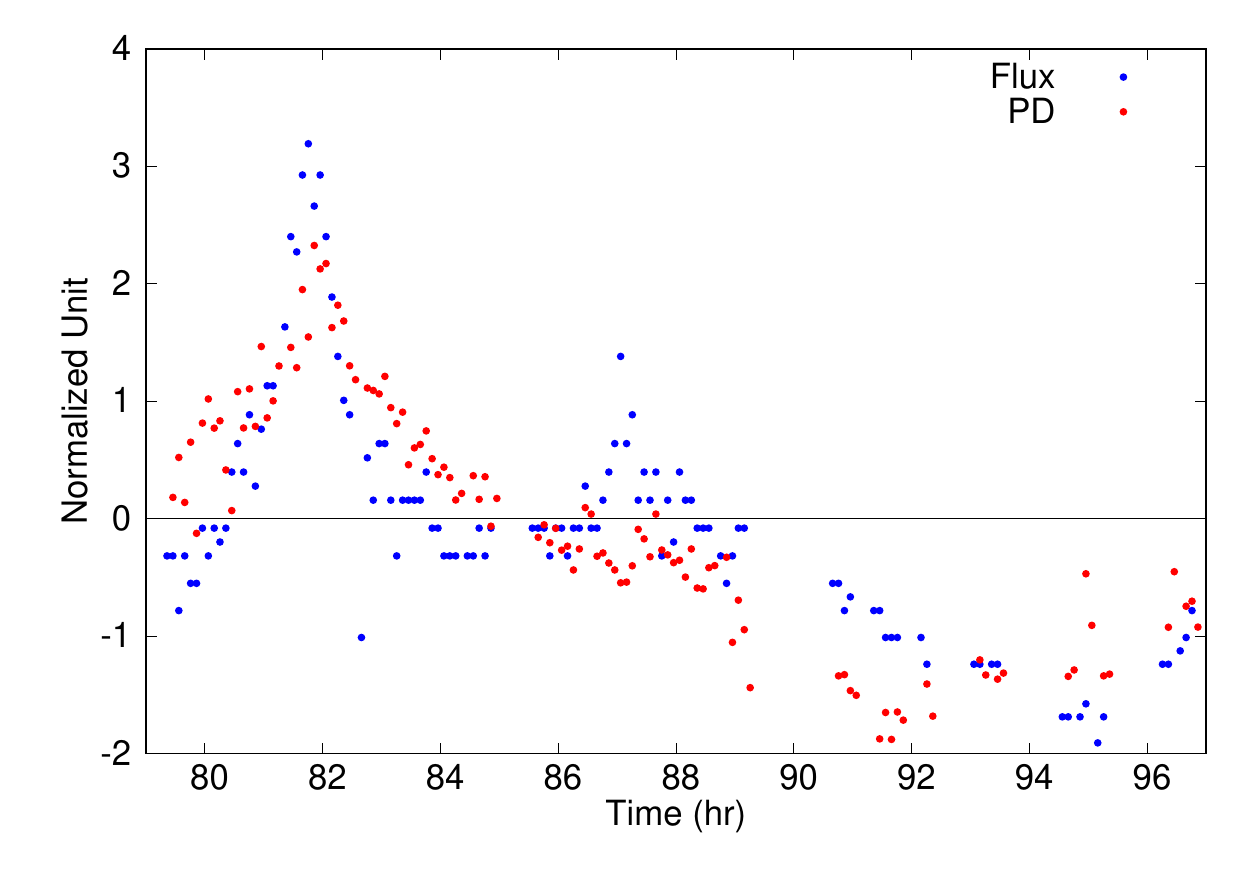} 
\includegraphics[width=\columnwidth]{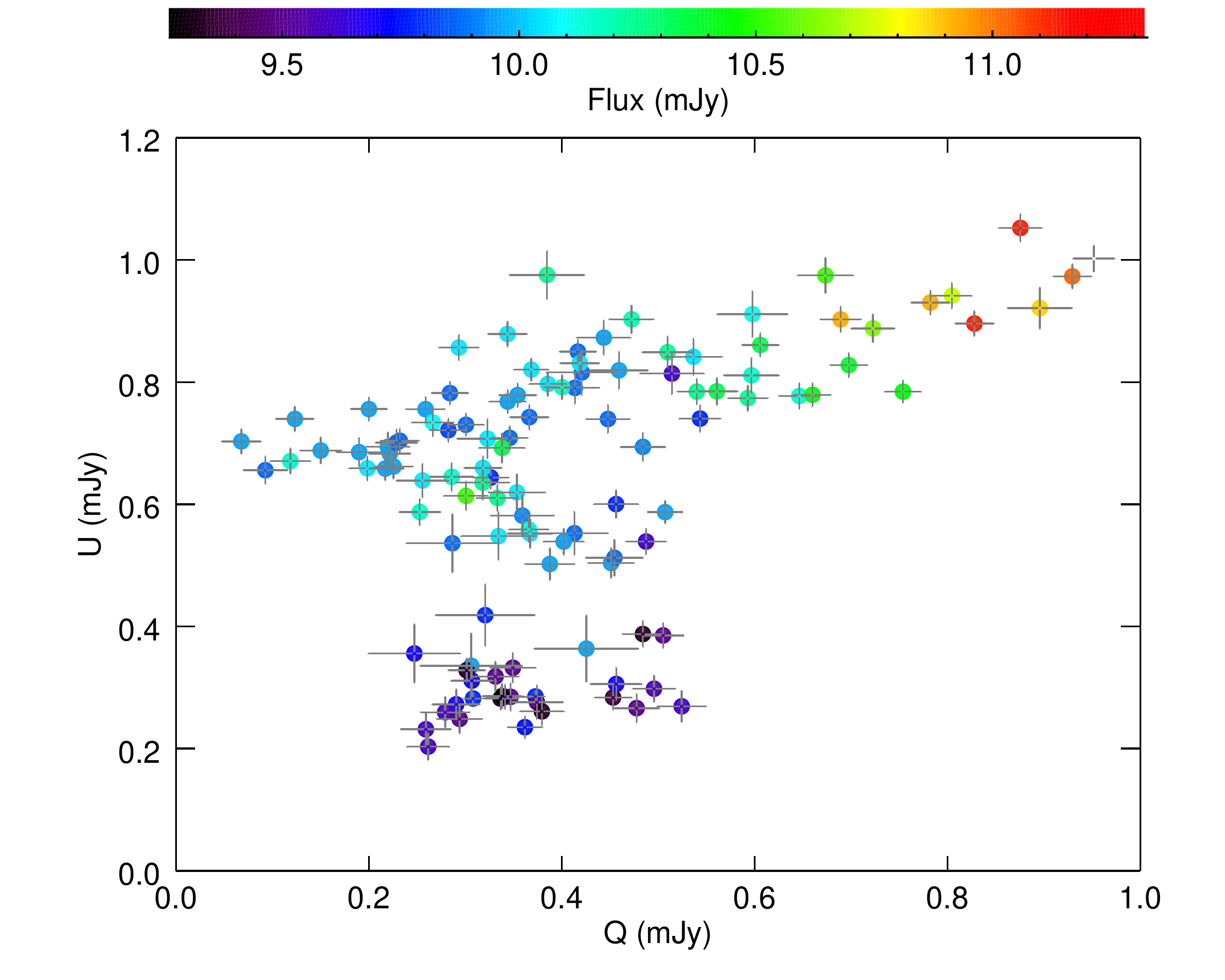}  
\caption{Same as Figure\,\ref{cor_1} but for Epoch\,II.}
\label{cor_2}
\end{figure}

For Epoch\,I, the calculated DCF between PD and the R flux is shown in the upper panel of Figure\,\ref{cor_1}. The analysis reveals a considerable high correlation (DCF value of $\sim 0.9$) with the 2\,h lag, such that the PD variations are leading flux changes. This lag can be seen clearly by eye even in the corresponding normalized light curves (mean subtracted and scaled by standard deviation) presented in the middle panel of the figure.  The correlation between PD and PA, on the other hand, was explored through the correlation between Stokes parameters $Q$ and $U$. A source evolution on the $Q-U$  plane, given in the lower panel of figure, reveals however no obvious relation between the PD and PA changes during the analyzed time interval (although note the large error bars).

For Epoch\,II, on the other hand, a significant correlation with zero lag has been found between the R-band flux and PD, implying certain level of unison between the total and polarized flux changes, as shown in the upper and middle panels of Figures\,\ref{cor_2}. This time, interestingly, PA and PD changes seem more structured as well, as presented in the lower panel of the figure. In particular, for higher fluxes a linear trend between $Q$ and $U$ can be observed. 

\begin{figure*}[t!]
\begin{center}
\begin{tabular}{ccc}
\resizebox{55mm}{!}{\includegraphics[angle=-90]{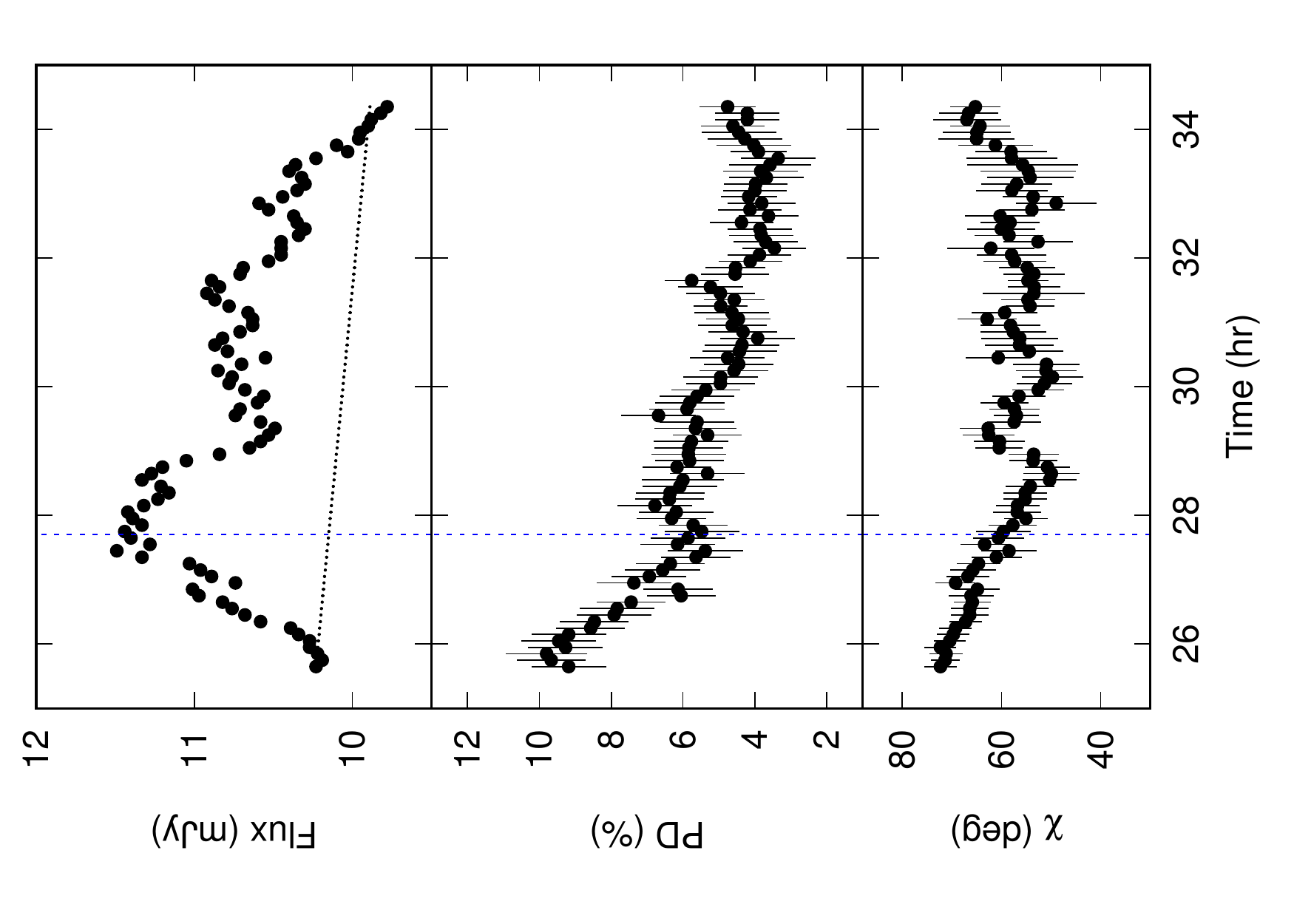}} &
\resizebox{55mm}{!}{\includegraphics[angle=-90]{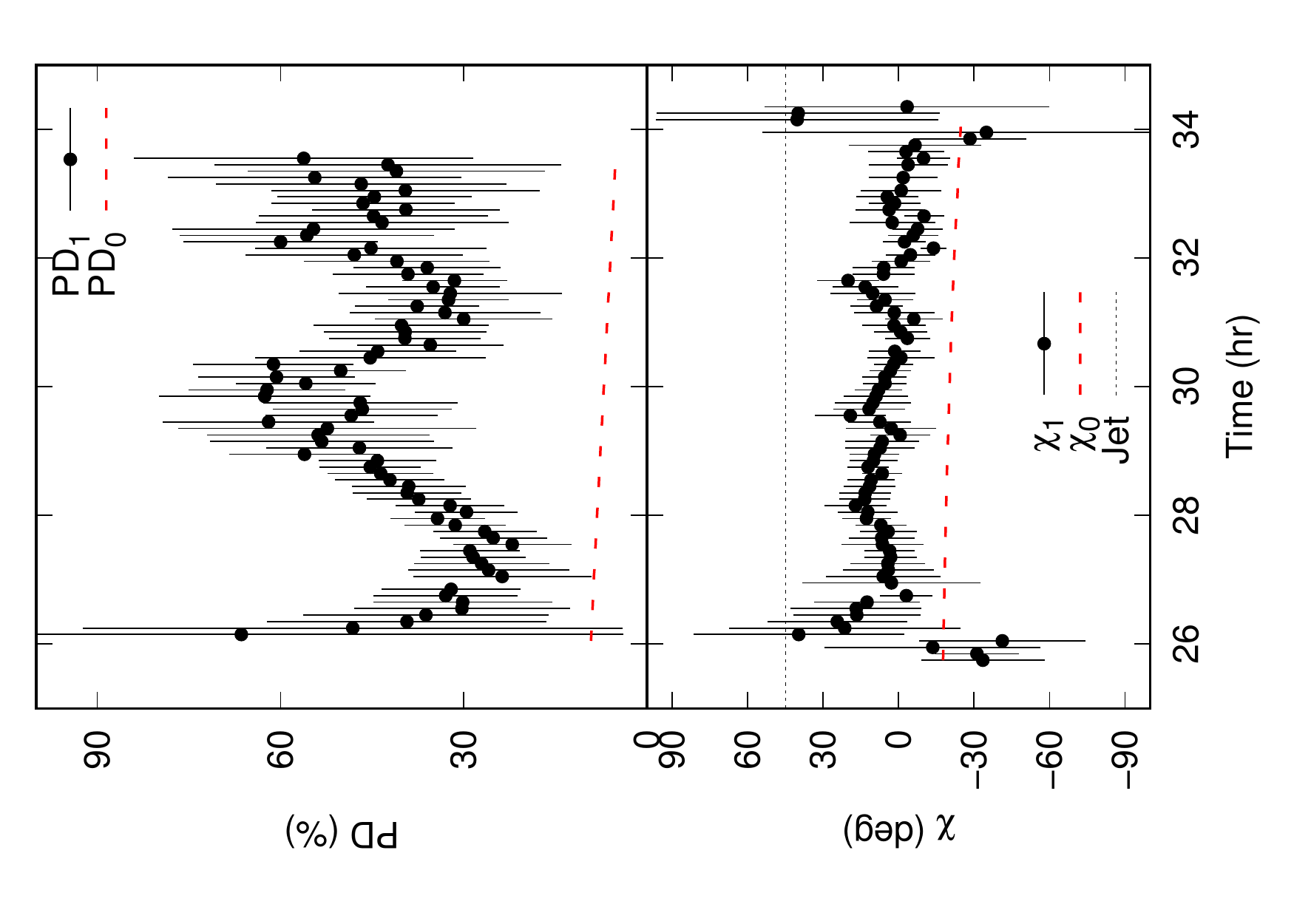}} &
\resizebox{55mm}{!}{\includegraphics[angle=-90]{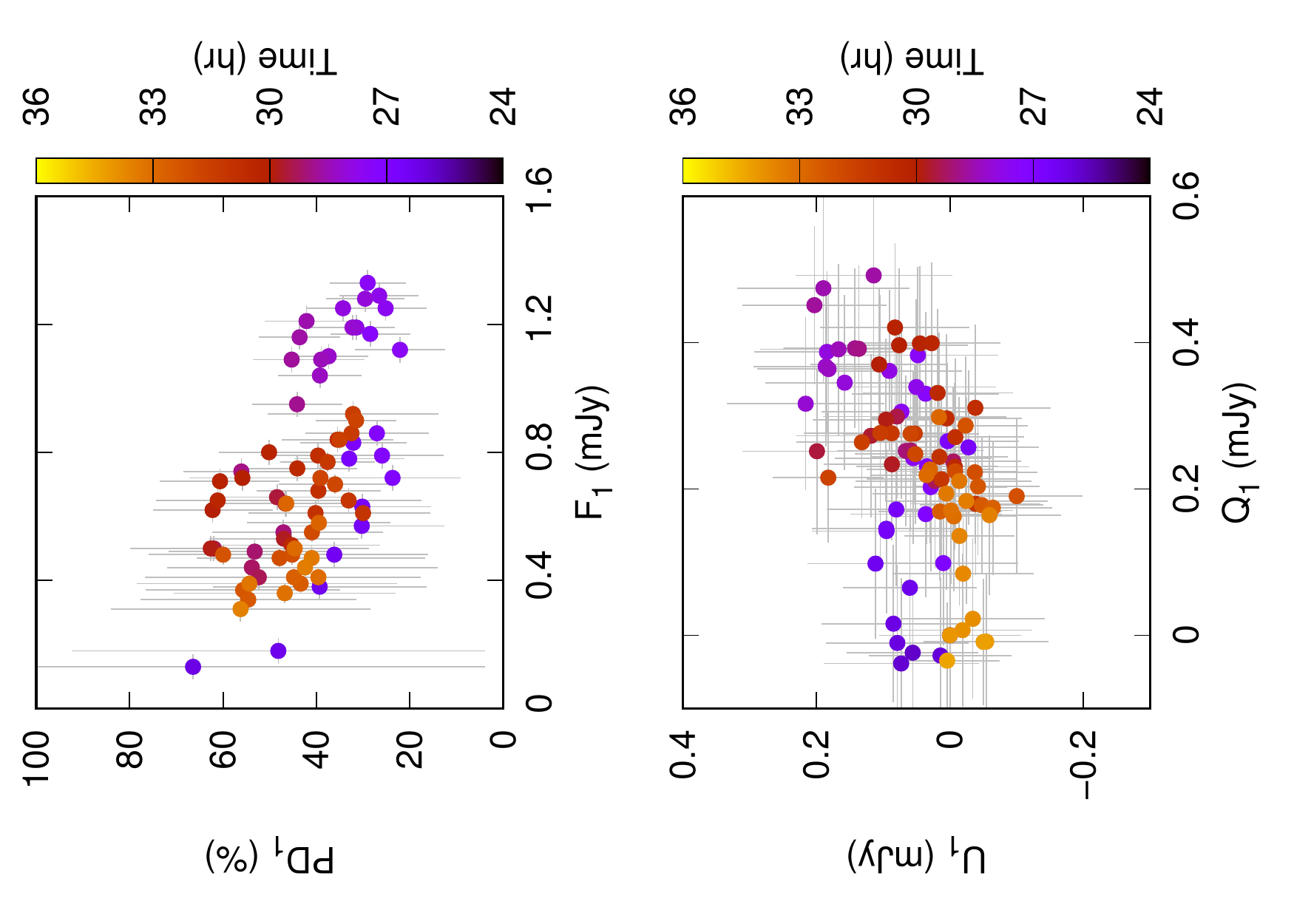}} \\
\end{tabular}
\caption{Photo-polarimetric analysis of the microflare\,1: The panels (from top to bottom) in the first column show the total flux, polarization degree, and polarization angle of the source in R band. In the second column, top and bottom panels present the polarization degree and polarization angle of the flaring ``pulse'' component, respectively, both subtracted from the slowly varying background component indicated in the plots by the dotted curves. The third column shows the variations in the microflare Stokes parameters $Q_1$ and $U_1$ (bottom panel), corresponding to the evolution on the $P\!D_1 - F_1$ plane (top panel). The vertical dotted line on the left column figure marks the segment of the light curve when the $P\!D$ clearly anticorrelates with the flux.}
\label{microflare1}
\end{center}
\end{figure*}

\begin{figure*}[t!]
\begin{center}
\begin{tabular}{ccc}
\resizebox{55mm}{!}{\includegraphics[angle=-90]{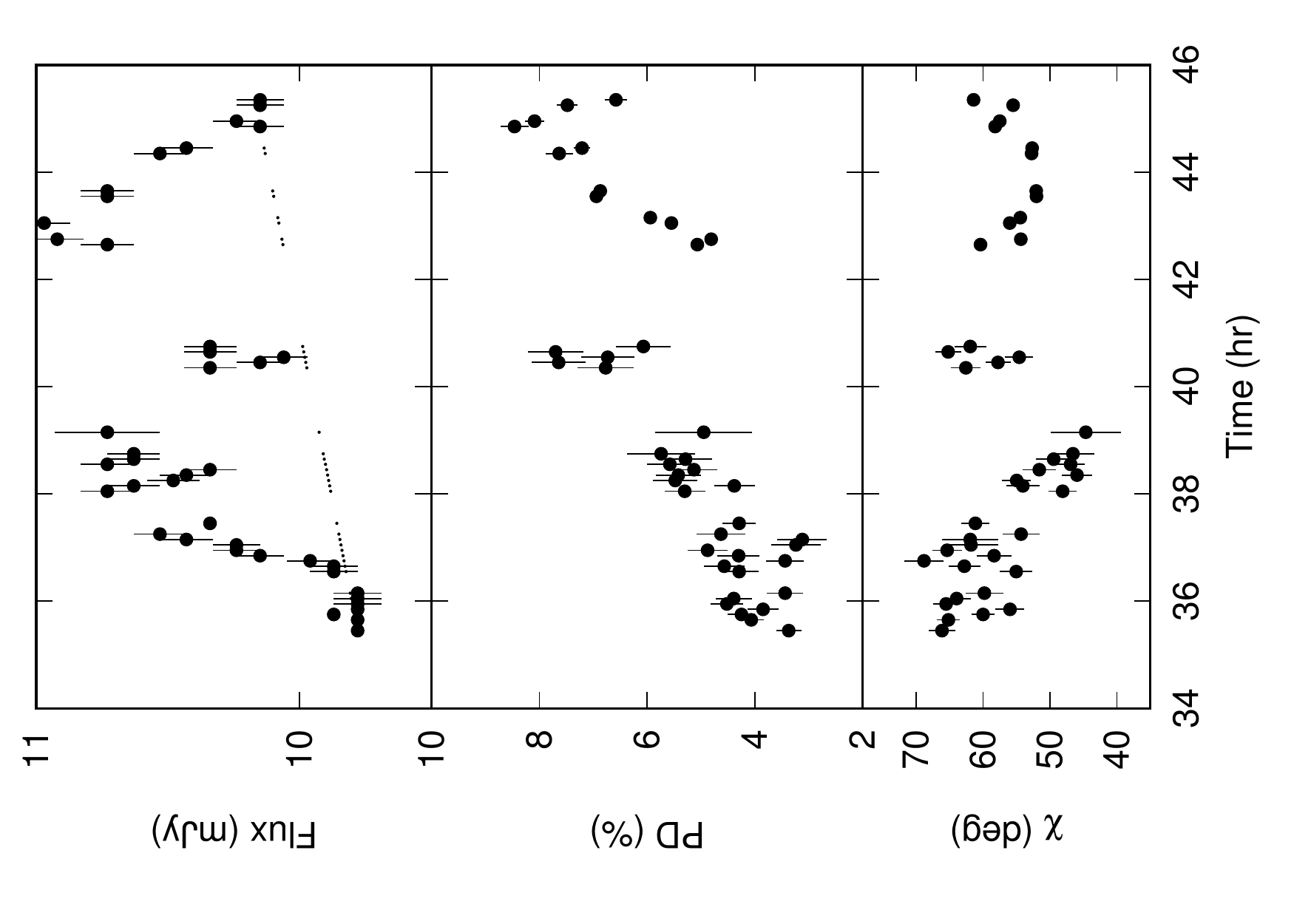}} &
\resizebox{55mm}{!}{\includegraphics[angle=-90]{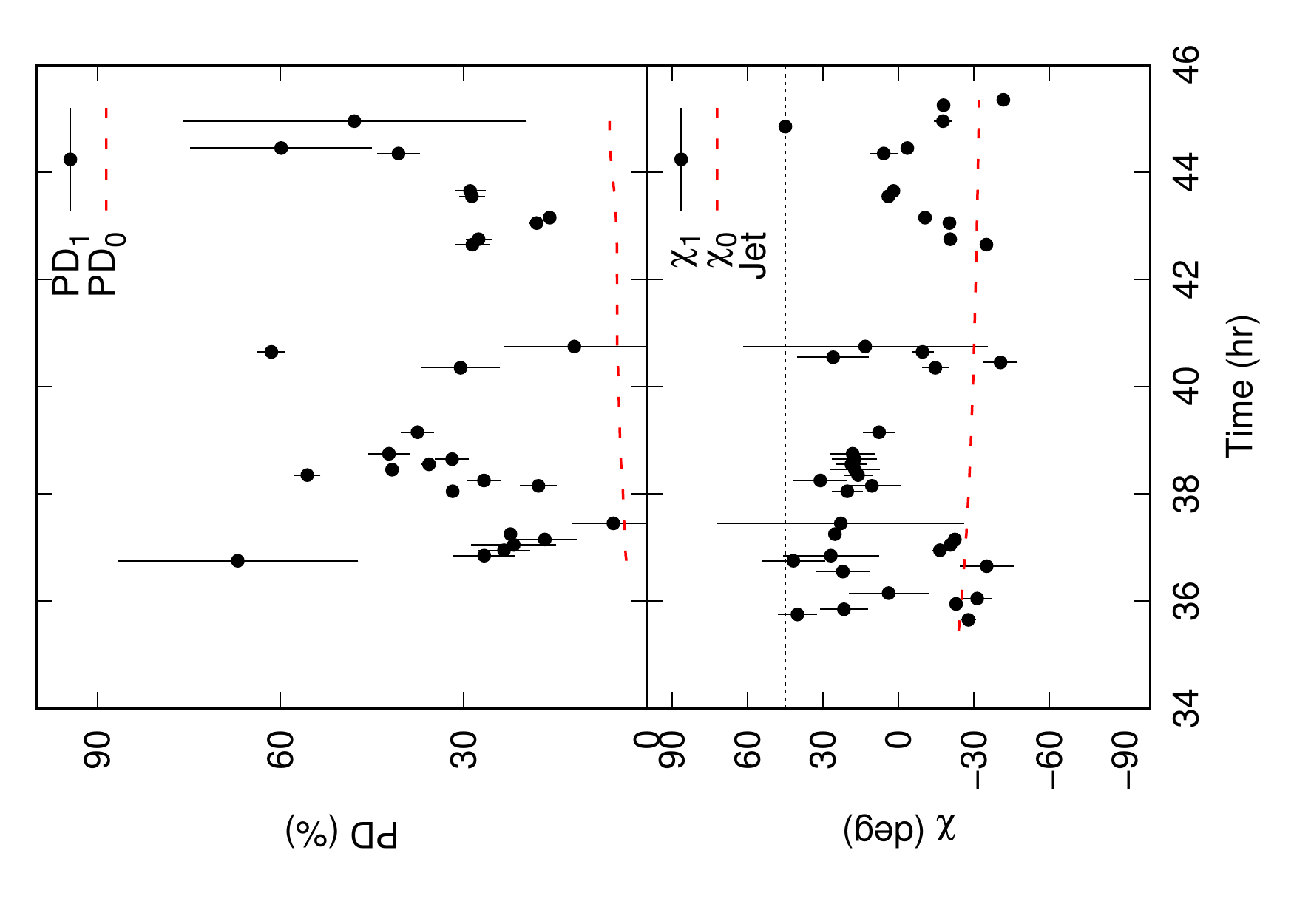}} &
\resizebox{55mm}{!}{\includegraphics[angle=-90]{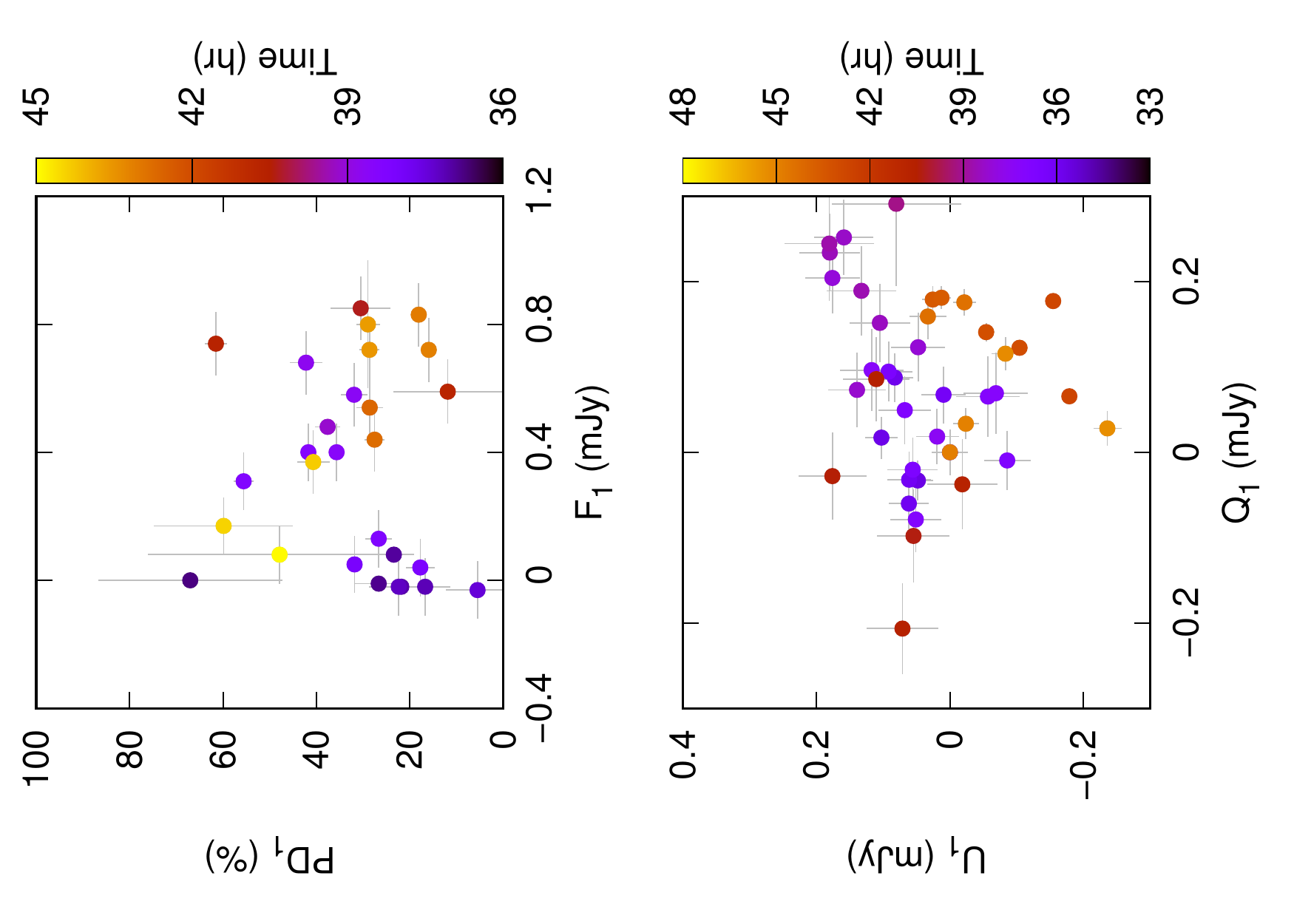}} \\
\end{tabular}
\caption{Same as Figure\,\ref{microflare1}, but for the microflare\,2.}
\label{microflare2}
\end{center}
\end{figure*}

\begin{figure*}[t!]
\begin{center}
\begin{tabular}{ccc}
\resizebox{55mm}{!}{\includegraphics[angle=-90]{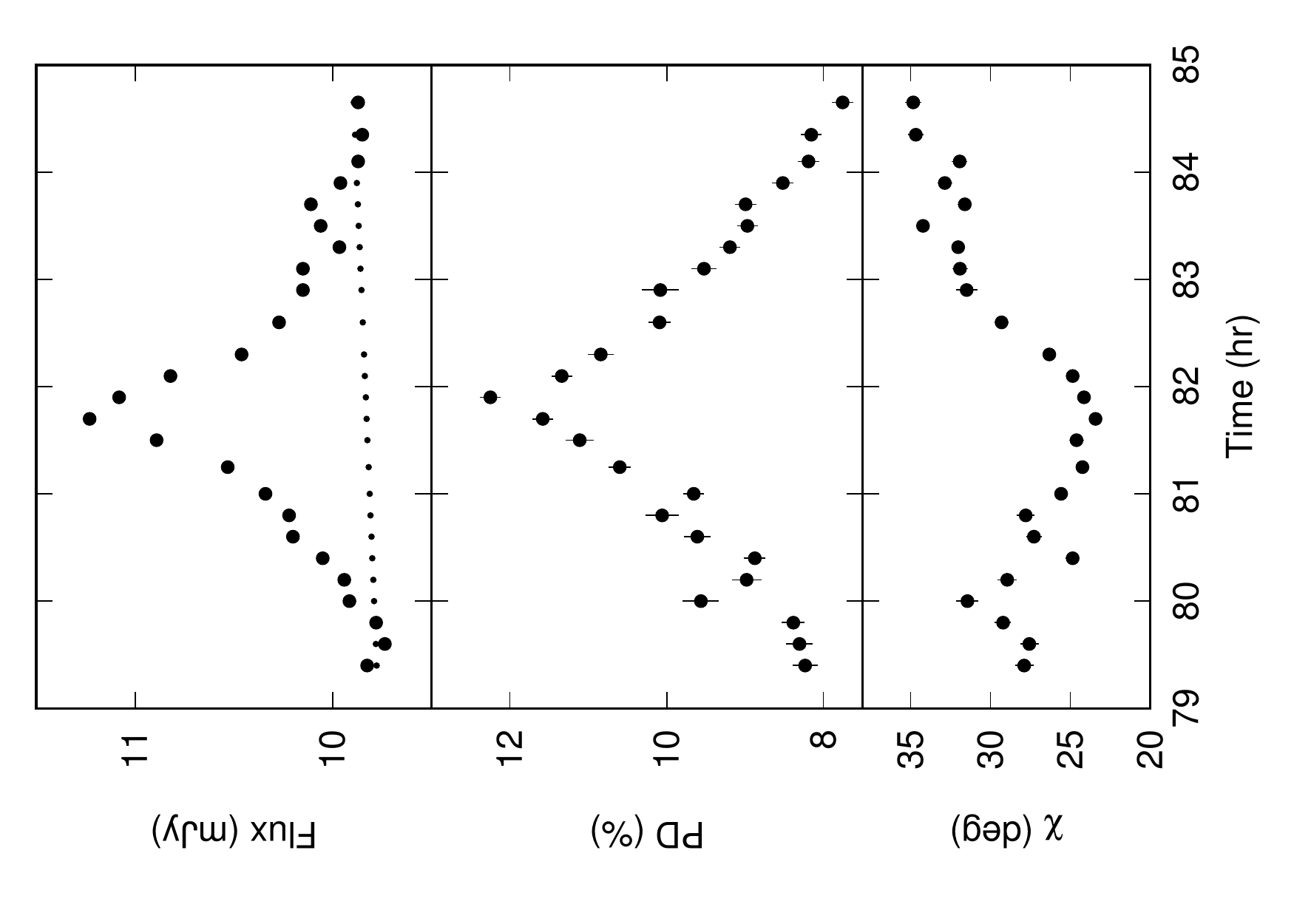}} &
\resizebox{55mm}{!}{\includegraphics[angle=-90]{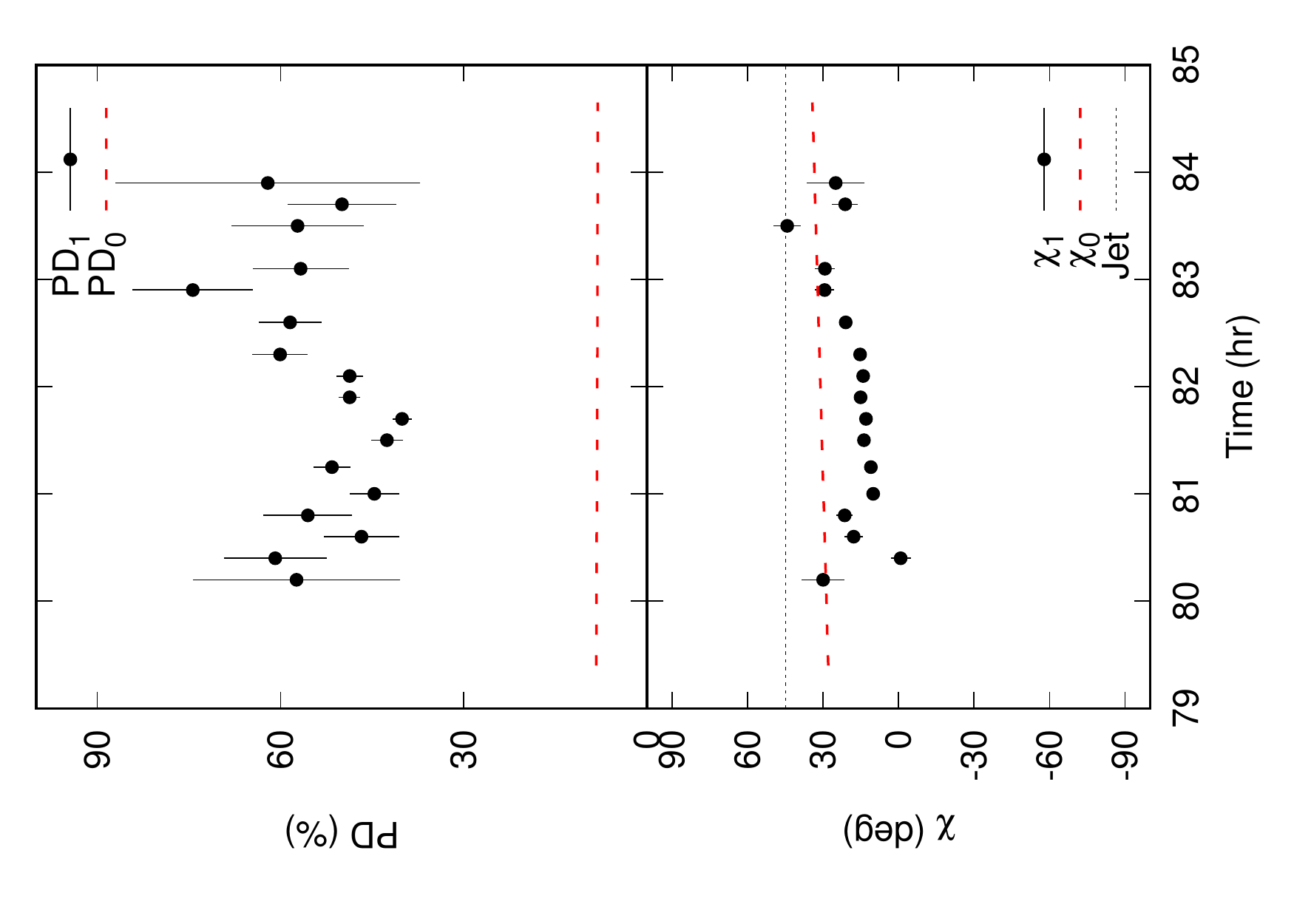}} &
\resizebox{55mm}{!}{\includegraphics[angle=-90]{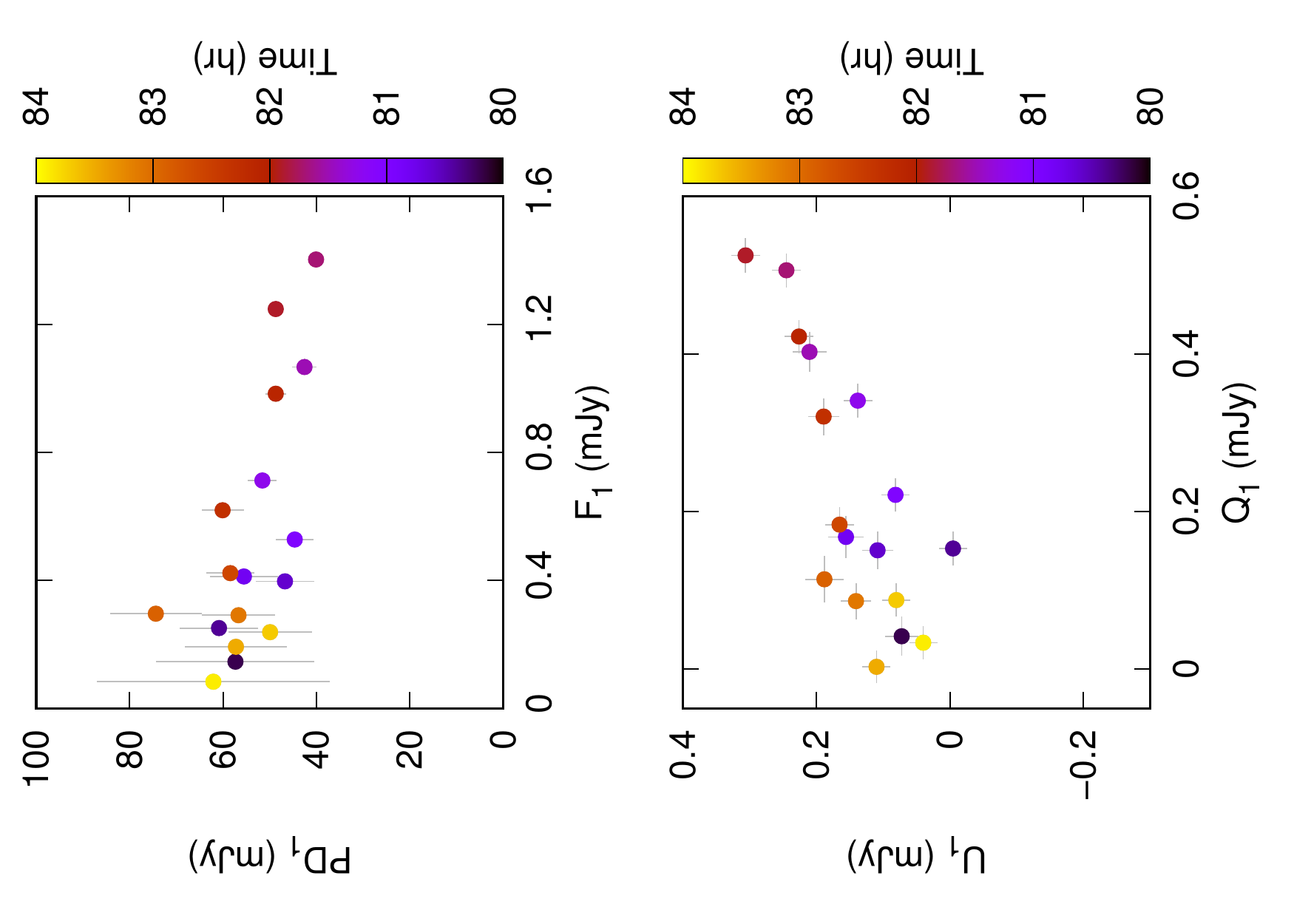}} \\
\end{tabular}
\caption{Same as Figure\,\ref{microflare1}, but for the microflare\,3.}
\label{microflare3}
\end{center}
\end{figure*}

\begin{figure*}[t!]
\begin{center}
\begin{tabular}{ccc}
\resizebox{55mm}{!}{\includegraphics[angle=-90]{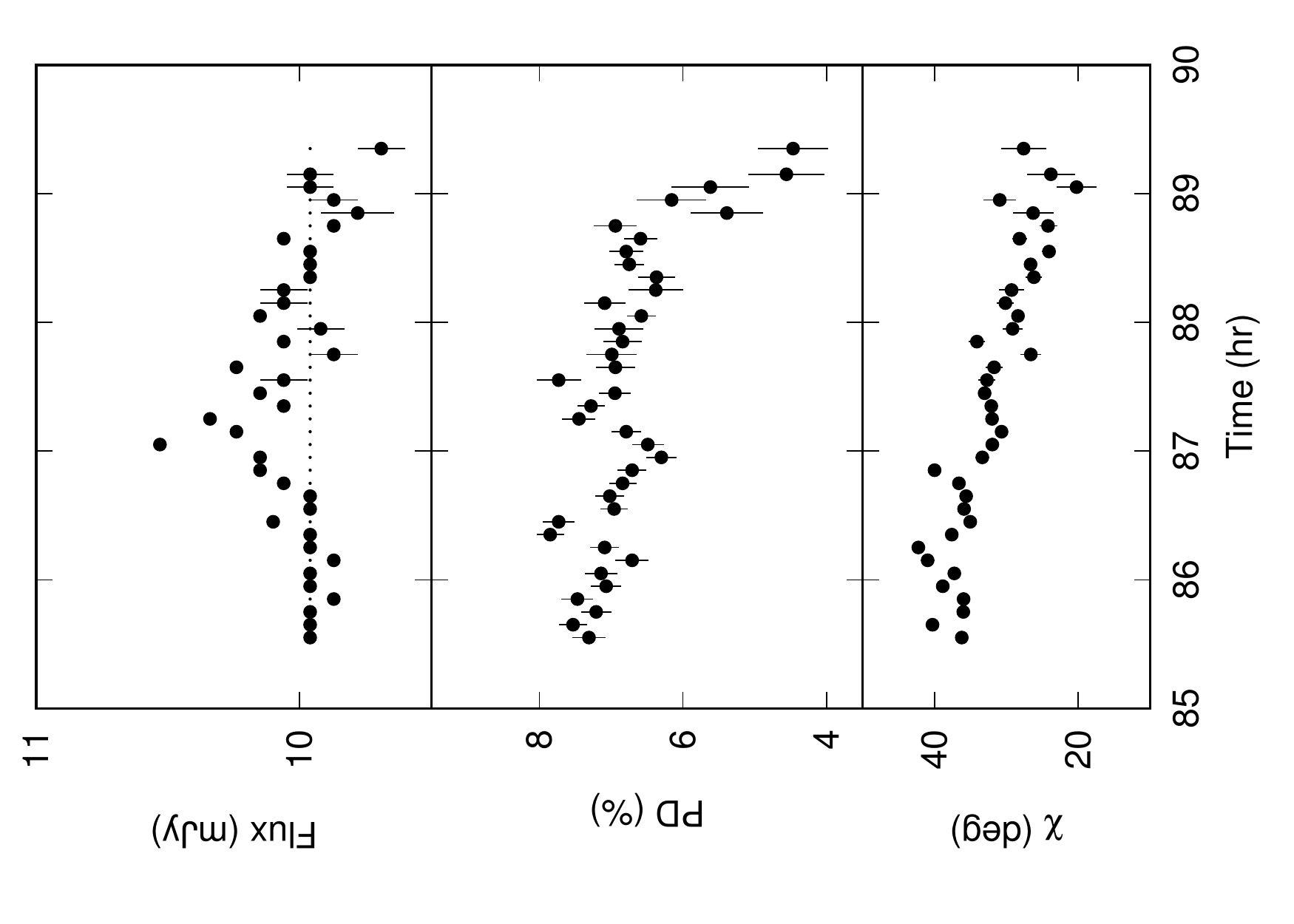}} &
\resizebox{55mm}{!}{\includegraphics[angle=-90]{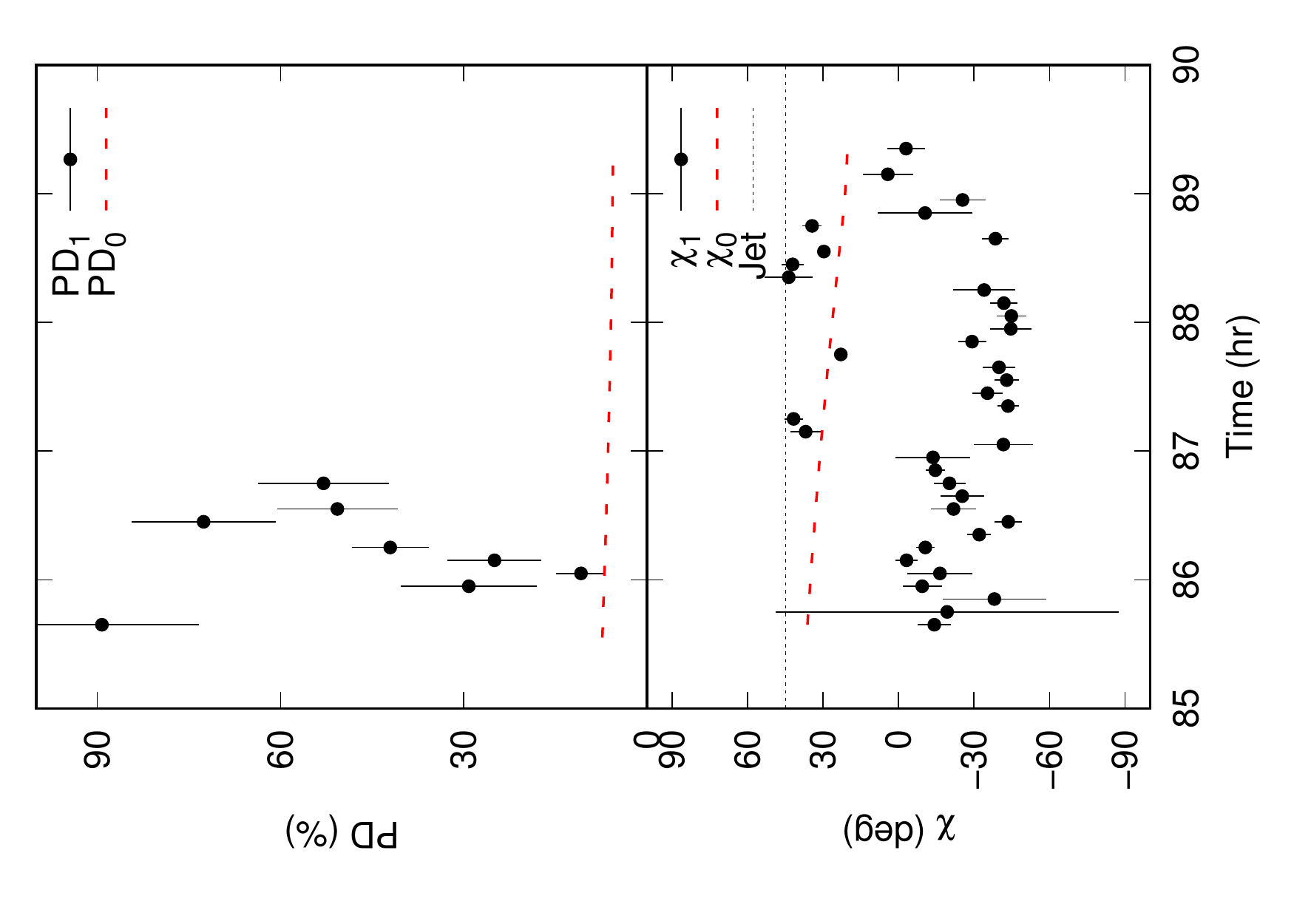}} &
\resizebox{55mm}{!}{\includegraphics[angle=-90]{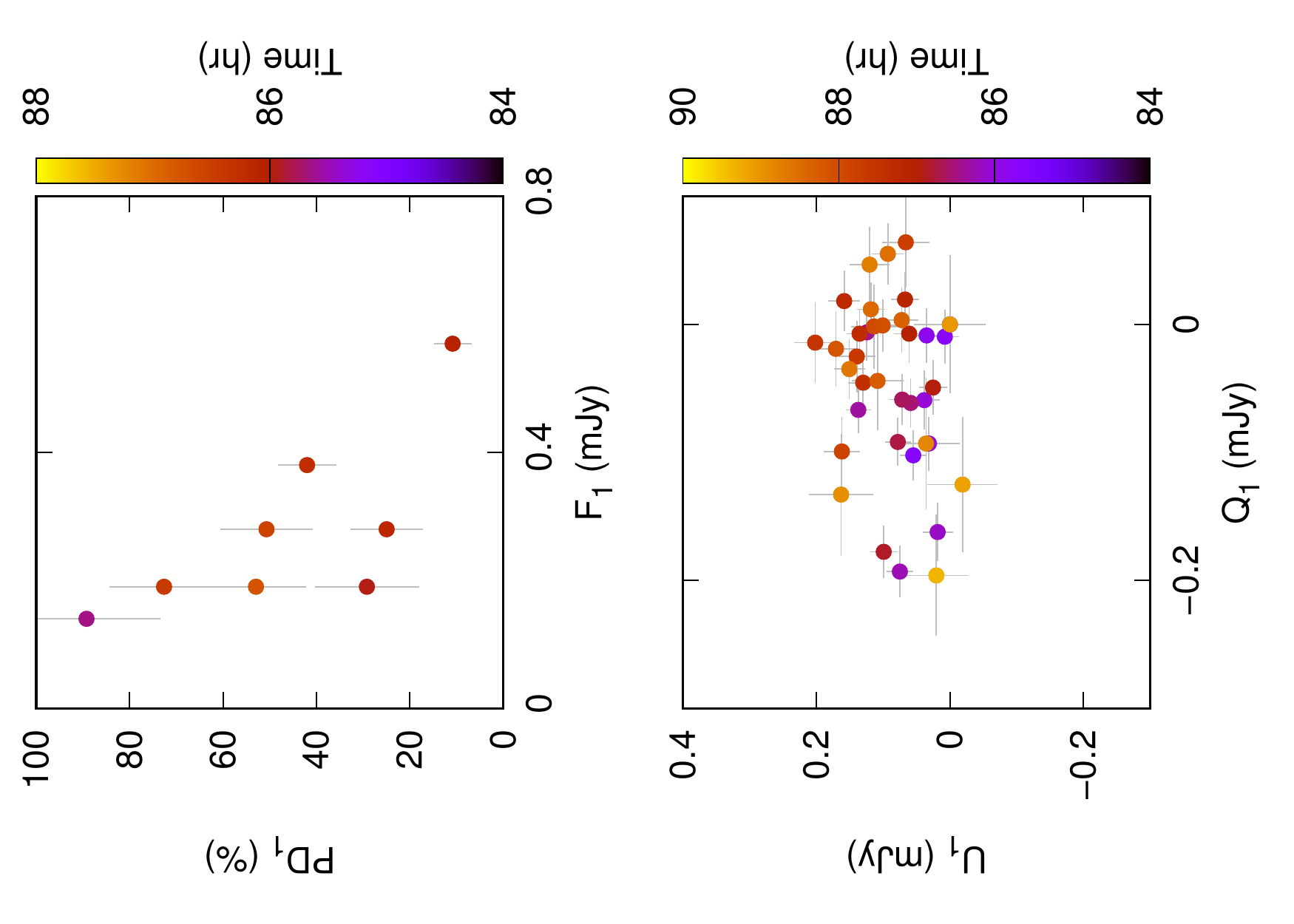}} \\
\end{tabular}
\caption{Same as Figure\,\ref{microflare1}, but for the microflare\,4.}
\label{microflare4}
\end{center}
\end{figure*}

\subsubsection{Modeling of Individual Microflares}
\label{sec:mod}

As shown in Figures\,\ref{rawlightcurve} and \ref{BVRI}, in addition to a day-long modulation of the S5~0716+714 light curve, we have detected also a number of rapid ``microflares'' during the 2014 WEBT campaign. Here we attempt to model some of them, assuming that they represent separate and distinct flaring events --- ``pulse emission'' components --- superimposed upon a relatively slowly-varying background component. In particular, making use of the simultaneous flux, PD, and PA measurements, for our analysis we have selected microflares detected during the time intervals 25--34, 34--46, 79--85, and 85--90\,h from the start of the campaign (marked in Figure \ref{rawlightcurve} by dashed vertical lines), which are shown in detail in the first columns of Figures\,\ref{microflare1}, \ref{microflare2}, \ref{microflare3}, and \ref{microflare4} (hereafter ``microflare\,1'', ``microflare\,2'', ``microflare\,3'', and ``microflare\,4'', respectively). An in-depth discussion on microflare\,3 is presented in \citet{Bhatta15}

Due to the linearly additive properties of total flux $F$ and the Stokes $Q$ and $U$ intensities, our base assumption regarding the distinctive nature of microflares implies
\begin{equation}
F = F_{0} + F_{1}  \, , \,\,\, Q = Q_{0} + Q_{1}  \, , \, {\rm and} \,\,\, U = U_{0} + U_{1}  \, ,
\label{stokes}
\end{equation}
where the ``microflare'' and the ``background'' emission components are denoted by indices ``1'' and ``0'', respectively. For each analyzed event, background intensities $F_0$, $Q_0$, and $U_0$ are estimated from fitting the data collected just before and just after a given microflare, and next microflaring intensities $F_1$, $Q_1$, and $U_1$ are found, giving us the microflare polarization degree $P\!D_1$ and polarization angle $\chi_1$
\begin{equation}
P\!D_1 = \frac{\sqrt{Q_1^2 + U_1^2}}{F_1}  \quad {\rm and} \quad \chi_1 = \frac{1}{2} \, \tan^{-1}\!\!\left(\frac{U_1}{Q_1}\right) 
\end{equation}
\citep[for further discussion see][]{Bhatta15}. The resulting evolutions in intensity and polarization of the selected events are presented in the second and third columns of Figures\,\ref{microflare1}--\ref{microflare4}. As shown, all the analyzed microflares are highly polarized, $P\!D_1 \geq 30\%$, but only microflare\,3 displays a clear looping behavior in $Q_1-U_1$ (or equivalently $P\!D_1 - F_1$) plain, with higher PD during the decaying phase of the pulse emission. Microflare\,1 exhibits a similar evolutionary pattern, with the overall anti-correlation between the flux and PD, but due to the large observational errors, any clear looping in the $Q_1-U_1$ plane can not be identified for this event with high confidence. Hints for the PD/flux anti-correlation can also be seen for microflares\,3 and 4. 

\begin{figure}[t!]
\centering
\includegraphics[width=\columnwidth]{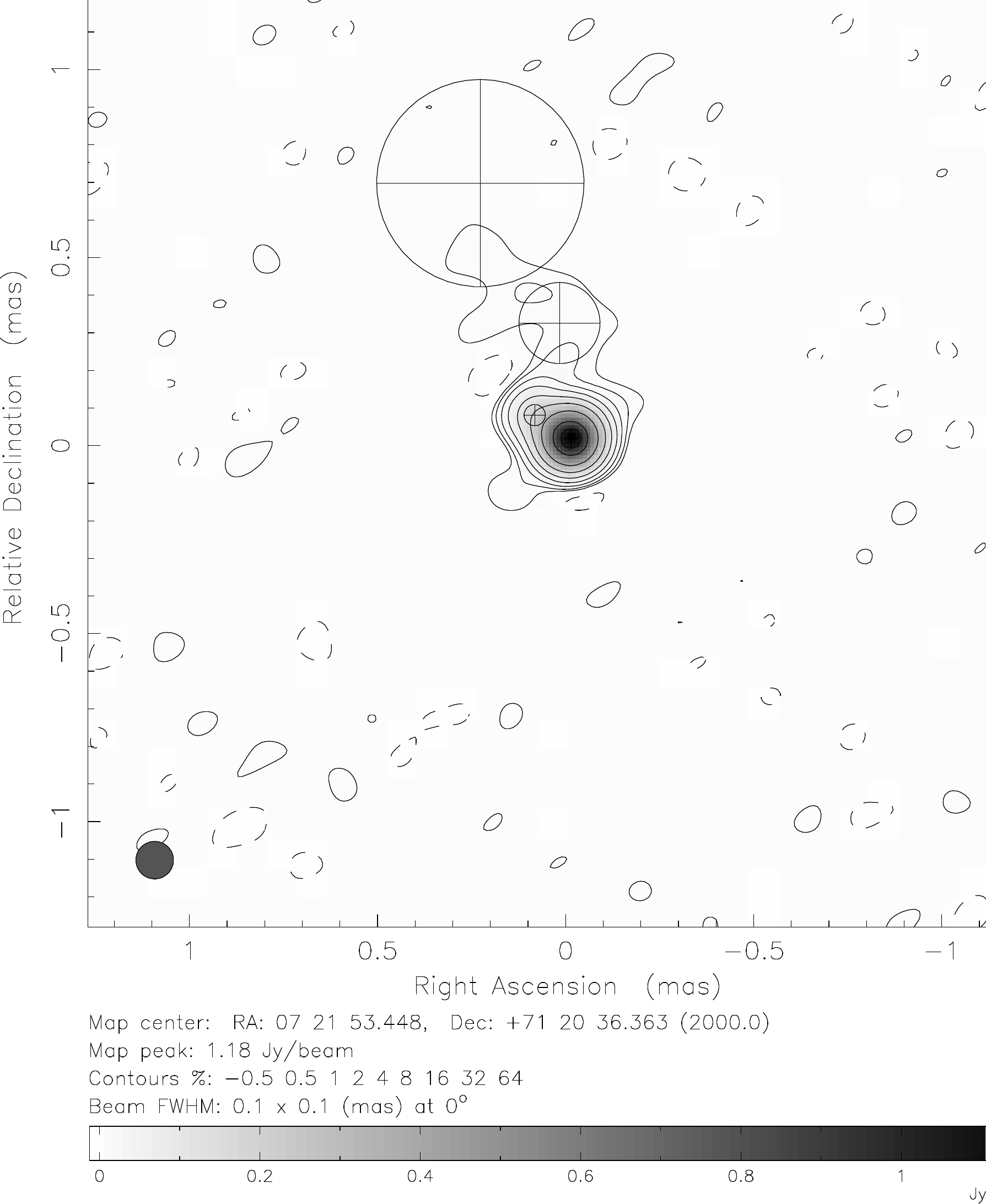}
\caption{Radio (VLBA-BU-BLAZAR) image of S5~0716+714 obtained at 43.135\,GHz on February 2014.}
\label{radio}
\end{figure}

An interesting difference between Epoch\,I and Epoch\,II can be noted here. Namely, while for the first two analyzed microflares\,1 \& 2 the PA of the pulse emission, $\chi_1 \sim 0-30$\,deg, is larger than that of the background components, $\chi_0 \sim -30$\,deg, being in addition relatively close to the jet position angle ($\sim 45$\,deg for the innermost parts of the outflow, i.e. within 0.12\,mas from the core, and $\sim 20$\,deg farther down the jet, according to the high-resolution radio image obtained on 2014 February 24 within the VLBA-BU-BLAZAR\footnote{\texttt{https://www.bu.edu/blazars/VLBA\_GLAST/0716.html/}} project; Figure\,\ref{radio}), for the latter two microflares\,3 \& 4 we derive $\chi_1 < \chi_0$ with $\chi_0 \sim 30$\,deg closely aligned with the jet axis.

\section{Discussion and conclusions}
\label{sec:con}

The 2014 WEBT campaign targeting S5~0716+714 was organized to monitor the source simultaneously in a number of the optical photo-polarimetric filters, for a longer period of time, in order to investigate in detail the evolution of flux, polarization degree, and polarization angle on timescales ranging from tens of minutes up to several days. The successfully conducted campaign, participated by many observatories all around the world, resulted in unprecedented dataset spanning $\sim 110$\,h of nearly continuous, multi-band observations (five consecutive days of flux measurements, including two sets of polarimetric data mainly in R filter, lasting each for about 25\,h with no major interruptions). The data were analyzed extensively using different statistical methods and approaches. The main observational findings can be summarized as follows:
\begin{itemize}
\item[1.] During the campaign, the source displayed a pronounced variability with peak-to-peak variations of $\gtrsim 30\%$, consisting of a day-timescale modulation with superimposed rapid (hourly-timescale) microflares characterized by flux changes by $\sim 0.1$\,mag; in general, variability amplitudes increase with the observing frequency.
\item[2.] The overall variability of the source is of the red noise type (consistent with a random-walk process); \emph{some} hints for the presence of quasi-periodic oscillations with the characteristic timescales of $3$\,h and $5$\,h have been found, but the in-depth analysis we have performed regarding these features, including an estimate of a ``global'' confidence bound in the source periodogram, as well as data folding, reveals that they do not represent highly significant departures from a pure red-noise power spectrum.
\item[3.] Flux changes in different bands track each other well, with no significant evidence for any time lags.
\item[4.] ``Bluer-when-brighter'' trend has been found in the source light curve, in a sense that flux maxima appear in general bluer than flux minima, but no tight correlation between the source flux and color could be established.
\end{itemize}

These results are broadly consistent to what was found before for S5~0716+714, in particular regarding the bluer-when-brighter trend \citep{Ghisellini97,Wu07,Sasada08,Poon09,Dai13}, although we note at the same time that the previous claims regarding the inter-band variability time lags in the source have been often contradictory \citep[e.g.,][]{Villata00,Qian02,Poon09,Wu12,Zhang12}, and also that the previous searches for the source quasi-periodicity were rather inconclusive \citep{Gupta08,Gupta09,Gupta12}. 

We argue that the bluer-when-brighter behavior implies that the observed flux enhancements are produced either during the episodes of an intensified particle acceleration, or alternatively by the fluctuating magnetic field superimposed on the underlying steady electron energy distribution with a concave shape. With respect to the source periodocity, we emphasize that the quality of the light curve analyzed here --- in particular its duration and uniquely dense sampling --- is basically unprecedented and as such perfectly suited for a search of hourlong quasi-periodic oscillations. The fact that we did not find such at the significance level high enough to claim the detection, is therefore very meaningful, implying no persistent periodic signal in the source within the analyzed variability timescale domain.

In addition to the above, the 2014 WEBT campaign resulted also in very novel, unexpected findings as well, namely:
\begin{itemize}
\item[5.] The $\sim 6$\,h-long period of the source inactivity has been observed; interestingly, in 2003 the blazar went through a very similar phase, at almost same ``quiescence/plateau'' flux level.
\item[6.] At a certain configuration of the optical polarization angle relative to the positional angle of the innermost radio jet in the source (Epoch\,I in \S\,\ref{sec:pol}), changes in the optical polarization degree led the total flux variability by about $2$\,h; meanwhile, at the time when the relative configuration of the polarization and jet angles altered (Epoch\,II), no time lag between polarization degree and flux changes could be noted.
\item[7.] The microflaring events, when analyzed as separate pulse emission components superimposed over a slowly-variable background, are characterized by a very high polarization degree ($>30\%$), and polarization angles which may differ substantially from the polarization angle of the underlying background component, or from the radio jet positional angle.
\end{itemize}

The peculiar plateau phase in the source light curve could be explained as resulting from a sudden but only temporary decrease in the jet production efficiency by the central accretion disk/supermassive black hole (SMBH) system. In this scenario, the observed optical emission of the blazar results from a superposition of fluxes produced within some larger portion of the outflow, from sub-parsec up to parsec scales, such that the emerging flux decreases with the distance, and the characteristic variability timescale increases (as a result of the jet radial expansion). A sudden disruption of the outflow at the jet base, resulting from some accretion disk instability around the jet launching region, would then result in a short-term ``disappearance'' of the highly variable innermost emission component, leaving only a slowly variable emission of the outer portions of the jet, and hence manifesting in the source light curve as a distinct plateau.

Note that the optical spectrum during the plateau phase is not much different from that observed during the rest of the 2014 WEBT campaign, indicating that the ``plateau flux'' is still due to the jet and not the accretion disk emission. Also, the fact that in 2003 a similar plateau has been observed at a similar flux level, which is however \emph{not} a historical flux minimum of the source, indicates that this outer emission component is not completely steady, but instead variable on very long timescales of years and decades. 

The 6\,h duration of the observed plateau could be linked to the characteristic timescale for re-building the outflow within the jet launching region, for which the \emph{shortest} one would be the Keplerian period around the innermost stable circular orbit (ISCO) of the accretion disk,
\begin{equation}
\tau_{K} = \tau_g \, \left(\frac{r_{isco}}{r_g}\right)^{3/2} \simeq 500 \, \left(\frac{\mathcal{M}}{10^8 M_{\odot}}\right) \, \left(\frac{r_{isco}}{r_g}\right)^{3/2} \, {\rm s} \, ,
\end{equation}
where $\mathcal{M}$ is the black hole mass and $\tau_g = r_g/c = G \mathcal{M}/c^3$ is the gravitational radius light-crossing timescale \citep[see, e.g.,][]{Meier12}. Hence, the 6\,h interval (seen both in 2014 and also in 2003), would imply $\mathcal{M} \simeq 4 \times 10^9 M_{\odot}$ for the maximally spinning SMBH ($r_{isco} \simeq r_g$), the value which should be considered as a safe upper limit for the S5~0716+714 black hole mass, or $\mathcal{M} \simeq 3 \times 10^8 M_{\odot}$ assuming very low spin values ($r_{isco} \simeq 6 \, r_g$).

During the 2014 WEBT campaign, we have also witnessed a very complex relation between the total intensity and the polarization properties of S5~0716+714.  In particular, during one brief incidence lasting $\sim 2$ h, the observed flux was found to be in clear anti-correlation with the  polarization degree as marked in the left column figure of Figure \ref{microflare1} \citep[see also][for the similar case in the blazar BL Lac in longer timescales]{Gaur14}; whereas considering the whole epoch the changes in the polarization degree were found to be leading the flux changes by about 2\,h. This suggests a delay between a build-up of the magnetic flux within the dominant emission region, and the onset of an efficient particle acceleration that follows, a behavior which could be reconciled with the scenario in which magnetic reconnection processes play a major role in the jet energy dissipation \citep[see in this context the most recent discussion in][]{Yuan16}. Yet during the subsequent epoch the optical polarization degree was well correlated with the optical flux, in agreement to what could be expected from the simplest model of a shock propagating along the jet \citep[see, e.g.,][and reference therein]{Hagen-Thorn08}, so the overall picture may not be unique. Still, the difference between the two epochs involved also a difference in the optical polarization angle, and in particular in an alignment of the polarization angle relative to the jet axis. Hence, it is possible that delays between the magnetic field build-up and the onset of particle acceleration are universal, but can be spotted only in the cases of a particular magnetic field orientation with respect to the jet axis and the line of sight.

A further insight into the energy dissipation processes in S5~0716+714, and other similar blazars, is provided by polarization properties of the shortest time-scale and smaller-amplitude fluctuations of the source. Such fluctuations are, in general, believed to be produced within small, possibly independent sub-volumes of blazar jets, that could be identified with isolated turbulent cells, magnetic reconnection sites, their mini-outflows, or small-scale shocks induced by such within the main jet body \citep[see in this context, e.g.,][]{Narayan12,Bhatta13,Marscher14,Calafut15,Chen16}. Here we have shown that, when modeled as distinct pulses superimposed on a slowly varying background component \citep[see in this context also][]{Hagen-Thorn08,Sasada08,Sakimoto13,Morozova14,Covino15,Bhatta15}, such microflares are always highly polarized, but at the same time are characterized by very different polarization angles which may deviate substantially from the polarization angles of the underlying background emission. 

In \citet{Bhatta15} we noted that, if blazar microflares are due to small-scale but strong shock waves propagating within the outflow, and compressing efficiently a disordered small-scale jet magnetic field component, one may expect various microflares to be characterized by very different polarization degrees, due to the fact that the expected value of the polarization degree depends strongly on the combination of the shock bulk Lorentz factor and the angle between the shock normal and the line of sight: even small changes in both parameters may result in significant changes in polarization degree! Yet what we observe during the entire 2014 WEBT campaign is that despite vastly different polarization angles of the microflaring events, the degree of the polarization is always very high. This finding calls for an alternative interpretation of blazar microflares.

\begin{acknowledgements}
The authors acknowledge support from the Polish National Science Centre grants DEC-2012/04/A/ST9/00083 (G. Bhatta, {\L}. Stawarz, M. Ostrowski) and 2013/09/B/ST9/00599 (S. Zola). The research at Boston University  was funded in part by NASA Fermi Guest Investigator grant  NNX14AQ58G and Swift Guest Investigator grant NNX15AR34G. The VLBA is an instrument of the National Radio Astronomy Observatory. The National Radio Astronomy Observatory is a facility of the National Science Foundation, operated under cooperative agreement by Associated Universities, Inc. The PRISM camera at Lowell Observatory was developed by K.\ Janes et al. at BU and Lowell Observatory, with funding from the NSF, BU, and Lowell Observatory. St. Petersburg University team acknowledges support from Russian RFBR grant 15-02-00949 and St.Petersburg University research grant 6.38.335.2015. G. Damljanovic, O. Vince and M.D. Jovanovic gratefully acknowledge the  observing grant support from the Institute of Astronomy and Rozhen National Astronomical Observatory, Bulgaria Academy of Sciences. This work is a part of the projects No.176011 (Dynamics and kinematics of celestial bodies and systems), No.176004 (Stellar physics), and No.176021 (Visible and invisible matter in nearby galaxies: theory and observations) supported by the Ministry of Education, Science and Technological Development of the Republic of Serbia. The Abastumani team acknowledges financial support of the project FR/639/6-320/12 by the Shota Rustaveli National Science Foundation under contract 31/76.  Shao Ming would like to acknowledge the support  by the National Natural Science Foundation of China under grants No. 11203016, 11143012 and  by the Young Scholars Program at Shandong University, Weihai. The authors acknowledge Luisa Ostorero for sharing the data and information on the 2003 WEBT campaign targeting S5 0716+714.
\end{acknowledgements}

\end{document}